\documentclass{aa}  
\usepackage{longtable}
\usepackage{ltcaption}
\usepackage{booktabs}
\usepackage{graphicx}

\begin{document} 
 
\hyphenation{pro-per-ties li-mi-ta-tions MEGASTAR res-pec-ti-ve-ly scien-ti-fic ini-tia-ted cha-rac-te-ris-tics Carras-co pa-ra-me-ters stelar des-cri-be de-di-ca-ted at-mos-phe-ric re-la-ted re-la-tive re-fe-rence attri-bu-ta-ble ma-na-ge SIMBAD co-lumn ge-ne-ra-tes crea-ted }
\hyphenation {cons-truc-ted mo-dels en-han-cing effects de-li-vers va-lua-ble Te-les-co-pio se-con-da-ry se-pa-ra-te-ly con-si-de-ring avera-ge re-sol-ving }
\newcommand{\HRR}{\mbox{HR-R}}
\newcommand{\HRI}{\mbox{HR-I}}
\title{MEGARA stellar spectral library}

   \subtitle{Second release}

   \author{E. Carrasco \inst{1},  M. Lira-Franco 
          \inst{1},  M.L. Garc{\'\i}a-Vargas \inst{2}, M. Moll\'{a} \inst{2}, N. Cardiel \inst{3,} \inst{4}, A. Gil de Paz\inst{3,} \inst{4}, P. G\'{o}mez-Alvarez \inst{2}, S. R. Berlanas \inst{5,} \inst{6,} \inst{7}
          \and 
         I. Mill{\'a}n-Irigoyen \inst{8}
          }

   \institute {Instituto Nacional de Astrof{\'\i}sica, {\'O}ptica y Electr{\'o}nica (INAOE)\\
   Calle Luis Enrique Erro 1, C.P. 72840, Santa Mar{\'\i}a Tonantzintla, Puebla, Mexico\\
              \email{bec@inaoep.mx}
               \and
         FRACTAL S.L.N.E., Calle Tulip{\'a}n 2, portal 13, 1A, E-28231 Las Rozas de Madrid, Spain
        \and Dpto. de F{\'\i}sica de la Tierra y Astrof{\'\i}sica, Fac. CC. F{\'\i}sicas, Universidad Complutense de Madrid, Plaza de las Ciencias, 1, E-28040 Madrid, Spain \and Instituto de F{\'\i}sica de Part{\'\i}culas y del Cosmos, IPARCOS, Fac. CC. F{\'\i}sicas, Universidad Complutense de Madrid, Plaza de las Ciencias 1, E-28040 Madrid, Spain \and Centro de Astrobiología (CSIC-INTA), E-28692 Villanueva de la Cañada, Madrid, Spain \and  Dpto. de Astrofísica, Universidad de La Laguna, E-38205, La Laguna, Tenerife, Spain \and Instituto de Astrof{\'\i}sica de Canarias, E-38200, La Laguna, Tenerife, Spain \and Max-Planck Institut für Astrophysik, 85741, Garching, Germany
}
   \date{Received September 15, 1996; accepted March 16, 1997}


 \abstract
   {We present the second release of the MEGARA spectral library, MEGASTAR, which now includes all spectra collected during ten observing semesters at the Gran Telescopio CANARIAS. This new release supersedes  our first release and incorporates a substantial number of additional observations (2000 new spectra), obtained at high spectral resolution full width half maximum R(FWHM)$\sim$20000, in two wavelength ranges centred on H$\alpha$ 
   (6420\,--\,6790\,\AA) and on the \ion{Ca}{ii} triplet (8370\,-\,8885\,\AA).}
   {The aim of this paper is to introduce MEGASTAR DR2 to the community as its high-resolution spectra can serve as a valuable resource for numerous types of research. In particular, we will use MEGASTAR spectra to construct  simple stellar population blocks within the HR-PyPopStar evolutionary synthesis models.}
   {The stars were observed using the integral field spectroscopy mode of the instrument. We processed the data in a uniform way with the MEGARA data reduction pipeline. We estimated the stellar flux by adding  the spectra from 37 spaxels, centred on the spaxel with the highest flux in the integral field unit reconstructed image. This approach guarantees that the effective slit width, and therefore the spectral resolution, are the same for all spectra.}
   {The second MEGASTAR release consists of 2838 spectra corresponding to 1408 stars, providing a better coverage of the stellar parameter space than the first release. The spectra were acquired with an average continuum S/N of about 215.}
   {This second release meets the standards of a modern empirical library: it offers reliable calibrations, data free from slit effects, observations of a large number of stars, and provides  high spectral resolution to model both individual stellar clusters and entire galaxies observed with MEGARA.}

      
\authorrunning{MEGASTAR }
\maketitle

\section{Introduction}
The study of galaxy spectra is based on population synthesis models. These models use combinations of simple stellar populations (SSPs) to reconstruct the star-formation histories of galaxies. Numerous authors have derived integrated properties or spectral energy distributions (SEDs) for SSPs, for example  \citet{mhk91, gvd94, bc03, cid05, gonzalez05, maraston05,  coelho07, eldridge09,  maraston11, conroy13, gvargas13, coelho14, vaz16, eldridge17, fioc19, stanway20, Millan2021, maraston25} and \citet{Millan2025}, among others.  The predictions of SSP models depend on several ingredients, such as the adopted isochrones, stellar libraries, spectral coverage, and numerical techniques used, among other factors.

One of the main elements affecting the models is the spectral resolution provided by stellar spectral libraries. According to \citet{trager12}, an ideal stellar spectral library is composed of stellar spectra obtained over the same wavelength range and resolution. The stars in such libraries are classified according to the primary atmospheric parameters that shape their SEDs, namely the effective temperature (T$_\mathrm{eff}$), the logarithm of surface gravity ($\rm \log{g}$), and the metallicity [Fe/H]. To reliably reproduce synthetic spectra, the stellar spectral library must span a wide range of values for each. 

In \citet{gvargas2020}, we examined the crucial role of spectral libraries in SSP models, both the benefits and limitations associated with empirical and theoretical libraries,  and highlighted that in all cases instrument-related properties, such as noise, instrument-specific effects, spectral resolution, and even the data reduction pipeline, add an additional level of uncertainty to the observed spectra. A notable exception is provided by empirical libraries constructed from stellar observations obtained with the same instrument. Our goal in building the MEGASTAR library was to precisely employ the spectra  obtained with MEGARA\footnote{Acronym of Multi Espectr{\'o}grafo en GTC de Alta Resoluci{\'o}n para Astronom{\'\i}a} as input for the {\sc HR-PyPopStar} evolutionary synthesis models \citep[][]{Millan2021,Millan2025}.\footnote{\url{https://cdsarc.cds.unistra.fr/vizbin/cat/J/MNRAS/506/4781}\label{nota1}}{$^{,}$}\footnote{\url{https://www.fractal-es.com/PopStar}}

A brief reminder of the main characteristics of the instrument: MEGARA offers both integral field spectroscopy (IFS) and multi object spectroscopy (MOS) at the Gran Telescopio CANARIAS (GTC). In its IFS mode,  the instrument uses an integral field unit (IFU),  named LCB (Large Compact Bundle), which covers a field of view of 12.5\,$\times$\,11.3\,arcsec$^{2}$. 
Additionally, it incorporates eight mini-bundles, each made up of seven fibres, assigned to sky subtraction. In MOS mode, the system deploys 92 robotic positioners, each hosting a mini-bundle of seven fibres, to sample a 3.5\,$\times$\,3.5\,arcmin$^{2}$ field of view on the sky.  In both modes,  the spatial resolution is 0.62\, arcsec per fibre.  This size corresponds to the diameter of the circle circumscribing the hexagonal spaxel. 
 The spaxel projection is oversized with respect to the fibre core in order to optimise the  fibre-to-fibre flux uniformity.
 
The instrument provides low (LR), medium (MR), and high (HR) spectral resolutions, corresponding to R(full width half maximum, FWHM) values of 6000, 12000, and 20000, respectively. The low- and medium-resolution configurations span the optical range from 3650 to 9750\,\AA, using six- and ten-volume phase holographic gratings (VPHs), respectively. The high-resolution configurations are dedicated to the rest-frame wavelength of H${\alpha}$ (HR-R) and the \ion{Ca}{ii} triplet (HR-I) covering the intervals of 6420\,--\,6790\,\AA\ and 8370\,--\,8885\,\AA, respectively. A detailed description of the instrument and its scientific commissioning is given in \citet{carspie18}, \citet{gilspie18}, and \citet{dull19}.
 
 For MEGASTAR we chose the IFS mode and the high-resolution configurations because when the project was initiated, no theoretical or empirical stellar library  matched the spectral resolution, wavelength coverage,  and instrumental characteristics of the MEGARA high-resolution setups HR-R and HR-I. Moreover, no other instrument on a 10 m-class telescope offers MEGARA-like performance in terms of spectral and spatial resolution over the same wavelength ranges. In addition,  our aim is to derive the stellar parameters T$_\mathrm{eff}$, $\log \mathrm{g}$, and [Fe/H] using a uniform methodology for all stars in the library.

This article is the fourth contribution in a series of papers. In Paper I, \citet{gvargas2020}, we introduced  the library, the criteria used for target selection, the initial observation programme, carried out during the MEGARA commissioning phase, and presented the data reduction pipeline.  In Paper II, \citet{carrasco2021} we  presented the first MEGASTAR data release (DR1), comprising 838 spectra of  414 stars. In Paper\,III, \citet{molla23}, we used a DR1 sample of  MEGASTAR members with spectral types later than B2 to derive T$_\mathrm{eff}$, log\,g, and [Fe/H], together with their associated uncertainties. The stellar parameters were obtained by matching the observed spectra to a grid of theoretical models, after rectifying to the continuum, both sets of spectra. This data release (DR1.1) comprised  351 stars and their corresponding data products.

This manuscript presents the second data release (DR2) of MEGASTAR. Its structure is as follows: Sections 2 and 3  describe the DR2 sample and the observing strategy, respectively. Section 4 outlines the data reduction procedure. The new features of the library database are highlighted in Section 5. Section 6 details the characteristics of DR2, and the last section provides the  concluding remarks.

\begin{figure}
\includegraphics[width=0.7\columnwidth,angle=-90]
 {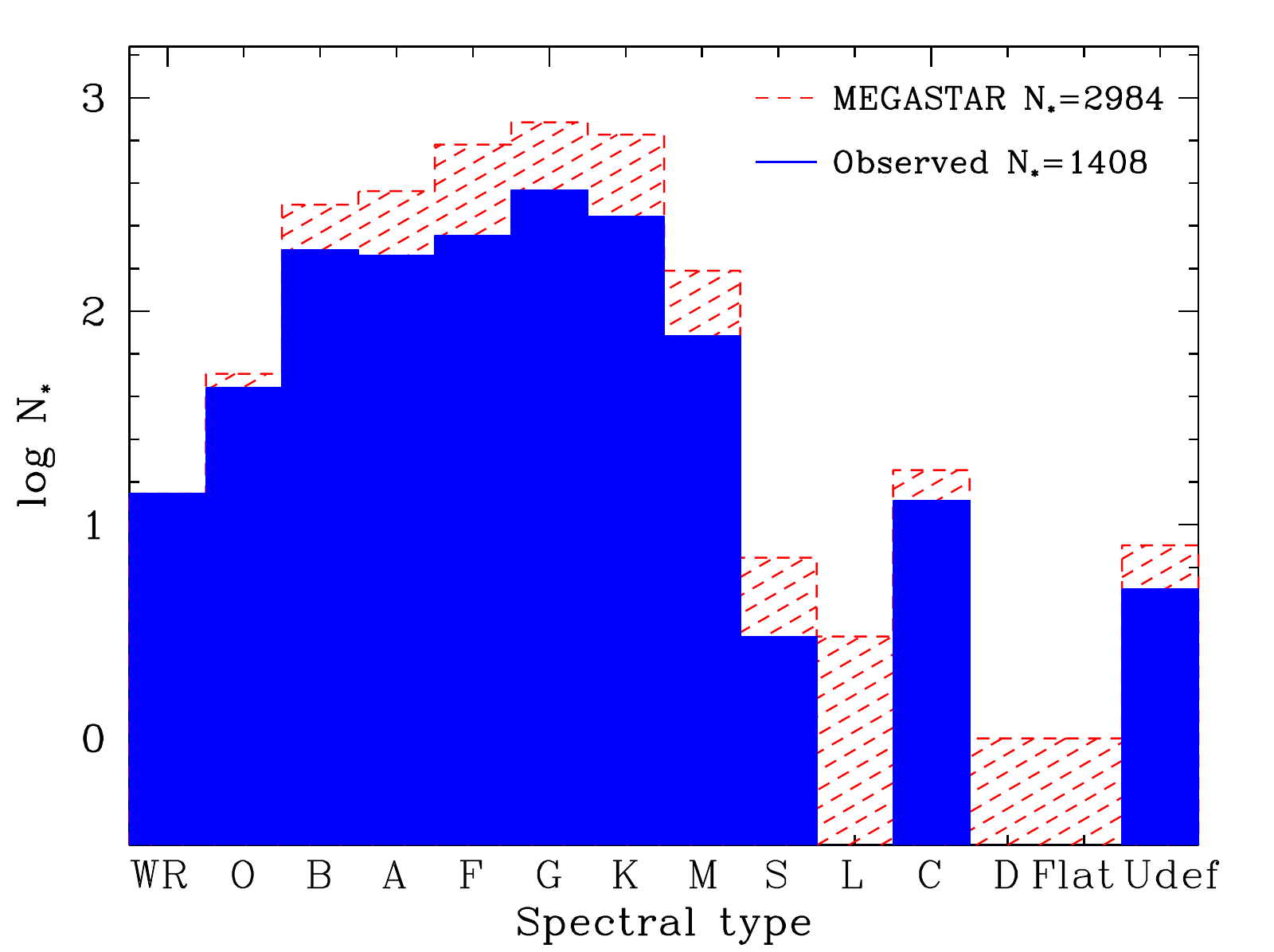}
 \caption{Distribution of MEGASTAR spectral types. Blue indicates those corresponding to the new DR2 data release, while red represents all stars in the complete library.}
\label{spectraltypes}
\end{figure}

\section{Sample}

We constructed the MEGASTAR catalogue, containing 2984 stars, by selecting targets from existing stellar libraries observed at spectral resolutions comparable to those of MEGARA and observable with the GTC. We also aimed for broad coverage in stellar parameters. After DR1, we obtained observations for 994 additional stars, re-observing six of them. As a result, DR2 consists of 2838 spectra from a total of 1408 stars.
 
\begin{figure}
\begin{center}
\includegraphics[width=0.9\columnwidth,angle=0]
{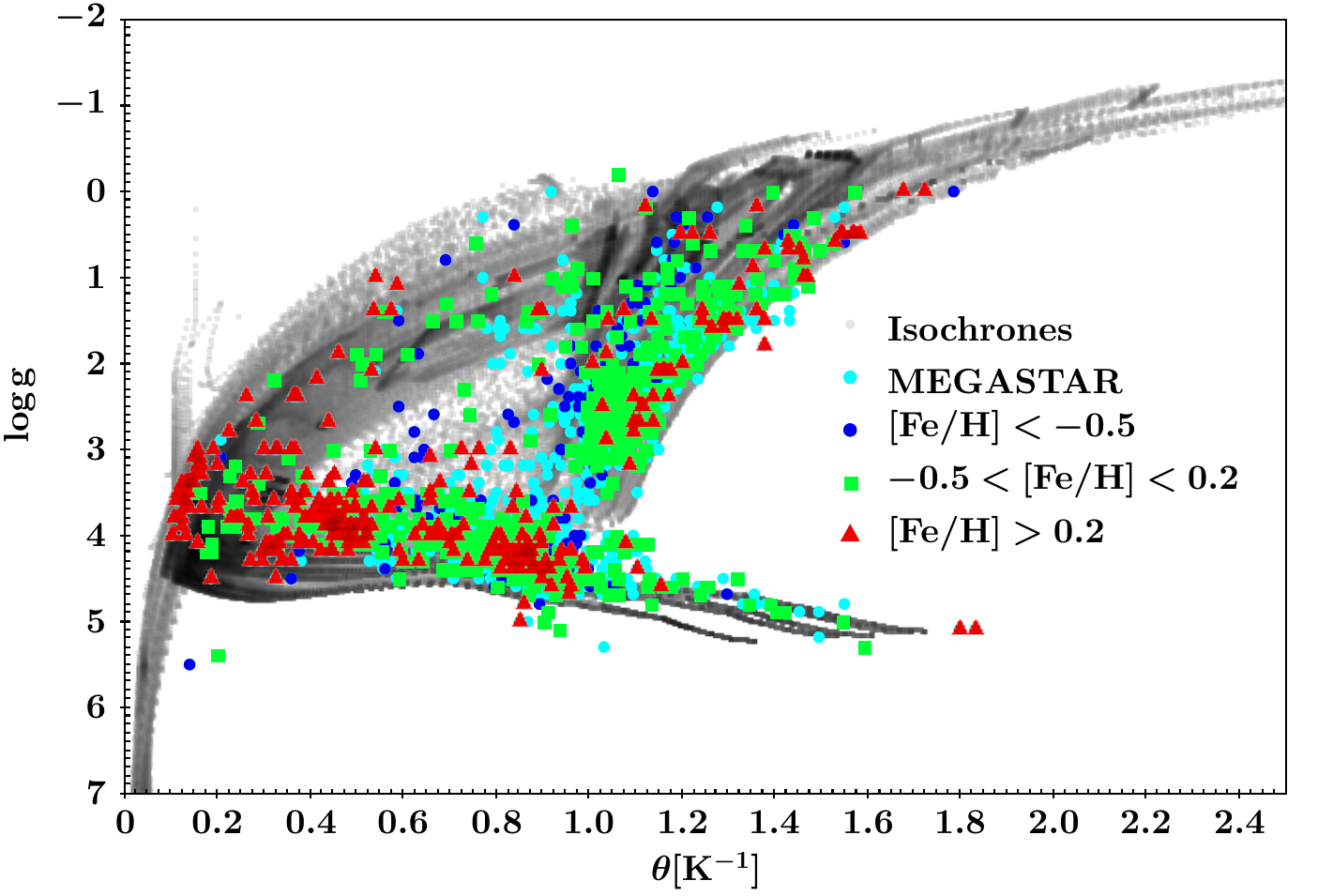}
\caption{Surface gravity  $\rm \log{g}$ {\sl vs.} $\Theta$=5040/$\rm T_{eff}$ diagram. We show: (1) the values given by the Padova isochrones (in grey) which are required for the synthesis code  {\sc HR-PyPopStar}, higher intensity corresponds to regions containing a larger number of points; (2) the library stars (in cyan) with stellar parameters taken from the literature; and (3) the DR2 stars divided into three metallicity ranges, each represented by different symbols and colours.}
\label{isochrones}
\end{center}
\end{figure}

Fig.~\ref{spectraltypes} presents the distribution of stellar spectral types compiled either from the SIMBAD-CDS database\footnote{\url{http://simbad.u-strasbg.fr/}} or from published sources. The most frequent spectral types in DR2 are G (370) and K (279), followed by F (227), B (194), A (183), M (76), O (44), W (14), C (13), and S (3). In Fig.~\ref{isochrones} the diagram of $\rm \log{g}$ {\sl vs.} $\Theta$  = 5040/$\rm T_{eff}$ is shown. We include Padova isochrones \citep[]{ber94, gir00, mar08}, which are used in the {\sc PopStar} evolutionary synthesis models \citep[][]{mol09} and in {\sc HR-PyPopStar} \citep{Millan2021, Millan2025}. This diagram enables us to identify areas of the physical parameter space that remain unobserved. We find a scarcity of metal-poor, cool giant stars, as well as late-type stars in general, primarily due to the characteristics of the filler-type observing programme, as discussed in Section \ref{observations}.

\begin{figure*}
\includegraphics[width=0.33\textwidth,angle=0]{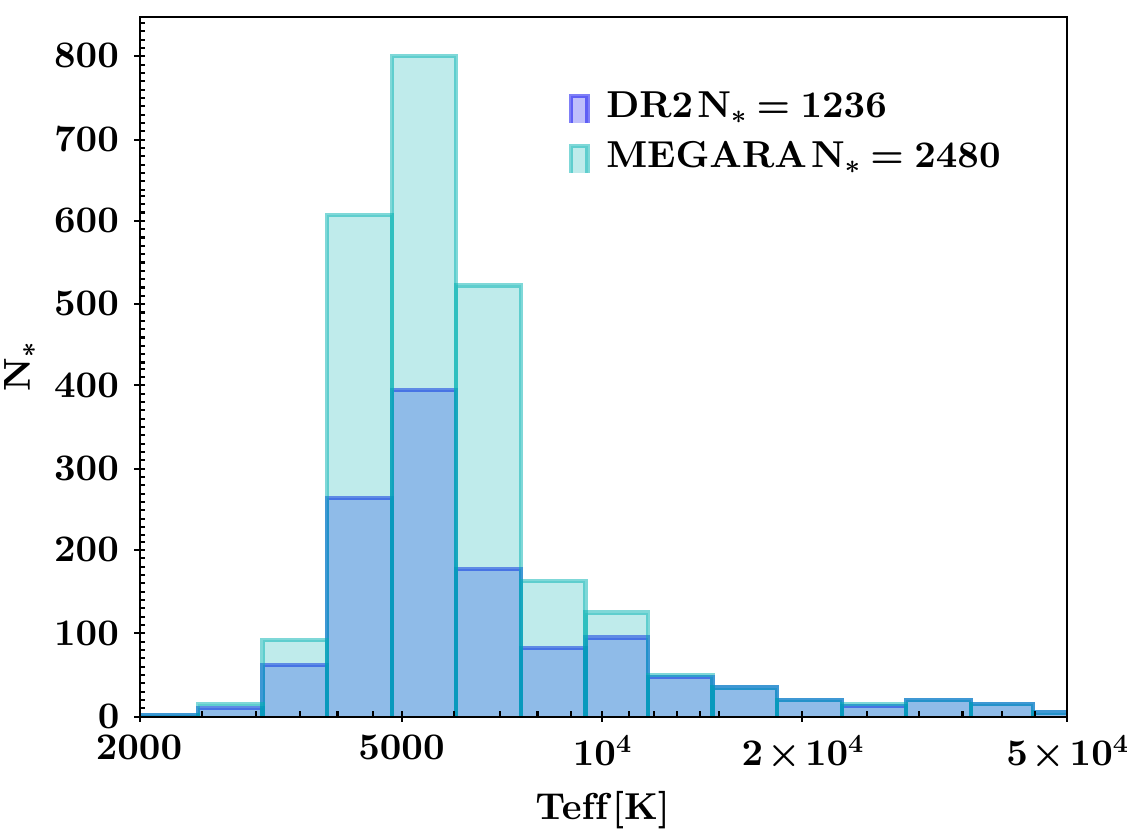}
\includegraphics[width=0.33\textwidth,angle=0]{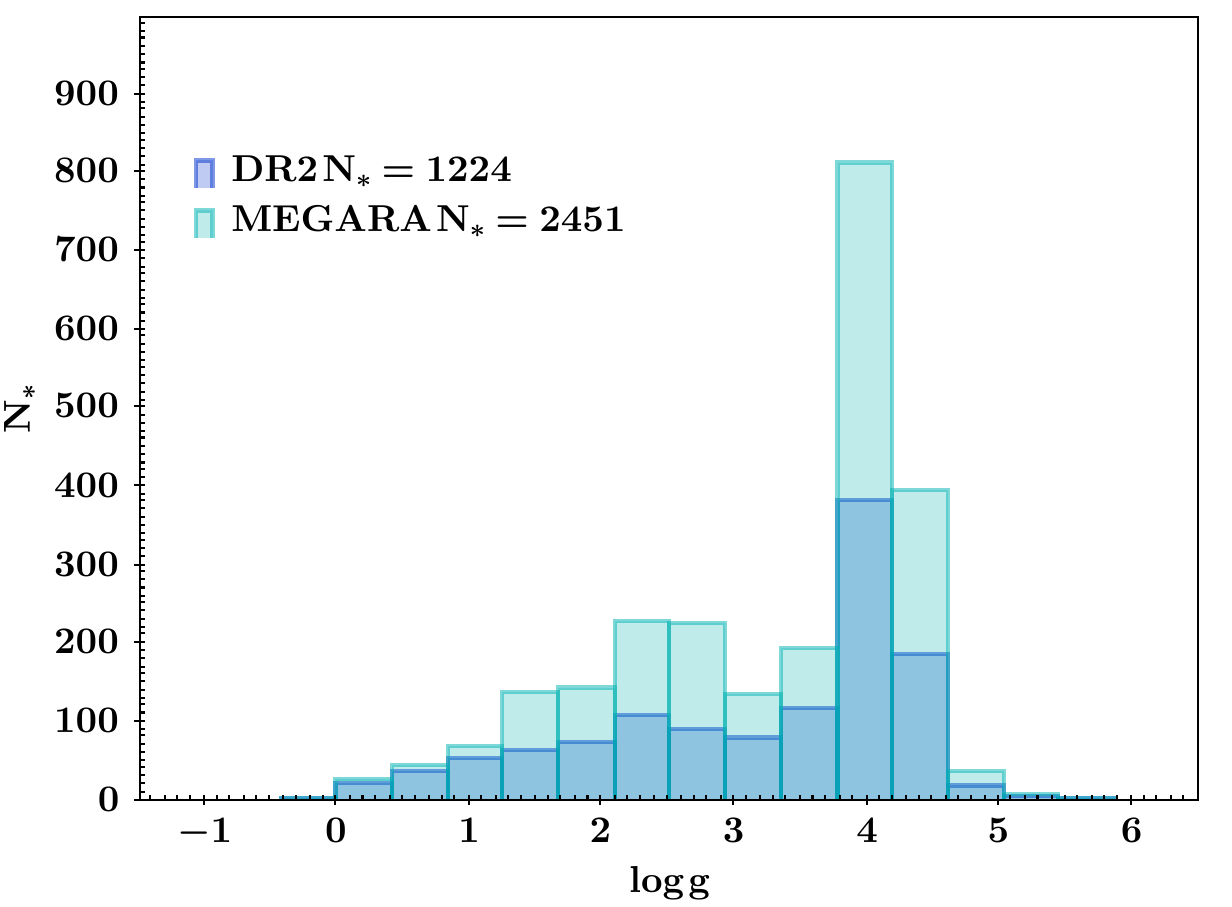}
\includegraphics[width=0.33\textwidth,angle=0]{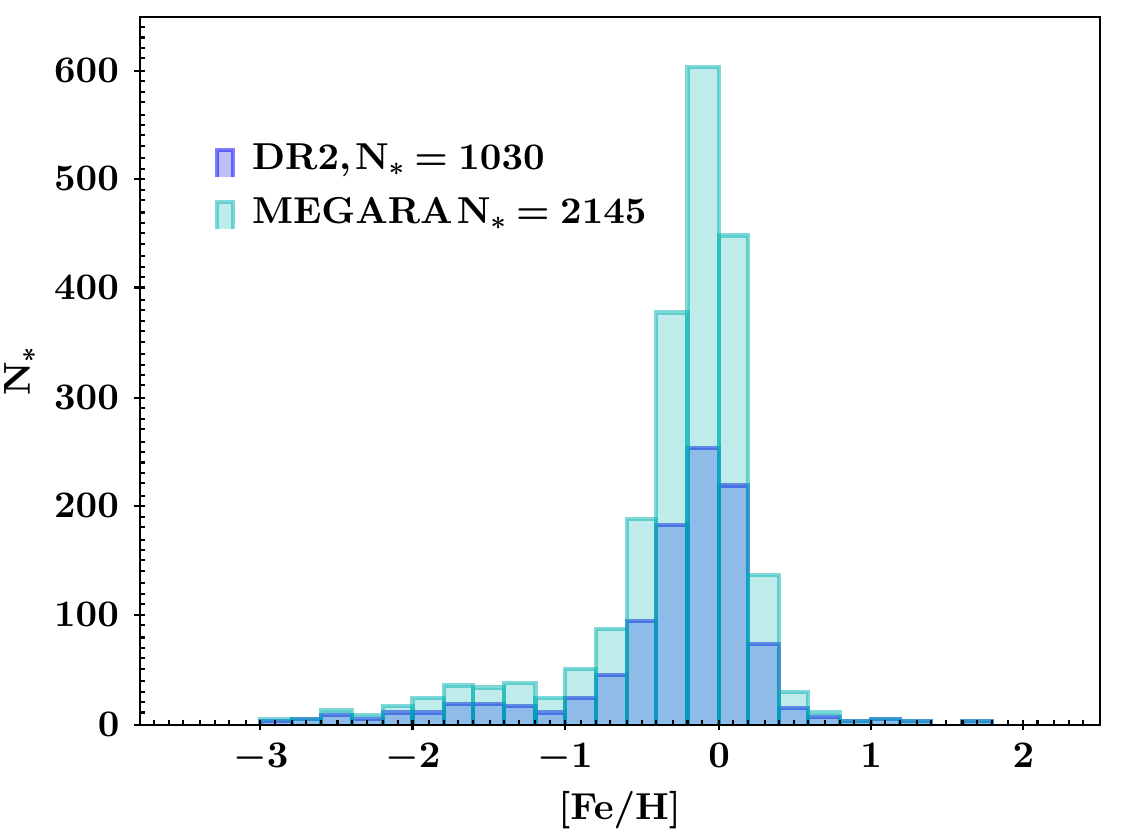}
\caption{Histogram of T$\rm_{eff}$ (left); log\,{g} (center) and  $\rm [Fe/H]$ (right) with values from the literature for  DR2 (blue) and MEGASTAR complete catalogue (green). The number of points is different in each graph as not all stars of the library have values of the stellar parameters from the literature.}
\label{histograms}
\end{figure*}

In Fig.\,\ref{histograms} we show the distributions of $\rm T_{eff}$, $\rm \log{g}$, and $\rm [Fe/H]$ for DR2 stars, compiled from the literature.  However, these stellar parameters are often dependent on the spectral resolution and wavelength coverage of the observations from which they were derived. 
A central goal of our project is therefore to determine the stellar parameters for all stars in the library in a homogeneous manner. For cool stars, we have already derived the parameters for the 351 objects with spectral types later than B2 in the DR1 sample (Paper\,III). In short, we used a minimisation method $\chi^2$ to match the observed spectra to the grid of theoretical stellar models of \citet{munari05}.

Hot, high-mass stars are analysed separately due to their inherent characteristics and associated effects (such as stellar winds, mass-loss rates, and often high rotational velocities, among others) that influence their spectra and necessitate dedicated stellar atmosphere models.
For OB stars, the stellar parameters are obtained using semi-automated tools designed to determine the physical properties of the stars, such as IACOB-broad and IACOB-gbat \citep[]{Simon-Diaz14, Simon-Diaz11}, based on extensive grids of synthetic spectra generated with the non-local thermodynamic equilibrium (LTE) FASTWIND stellar atmosphere code, which accounts for sphericity and mass loss \citep[]{Santolaya97, Puls05} and has been further adapted to cover the I-band. This model grid spans the broad interval of stellar and wind parameters adopted for standard OB-type stars, encompassing spectral types from early O to early B and luminosity classes from dwarfs to supergiants. Because WR star spectra are dominated by intense wind features, they will be studied with the non-LTE CMFGEN stellar atmosphere code, specifically developed for massive stars exhibiting strong winds and high mass-loss rates, such as Wolf-Rayet (WR) stars \citep{hillier01}. The corresponding results will be presented in the next paper of this series (Berlanas et al., in prep.).

\begin{figure}
\begin{center}
\includegraphics[width=0.40\textwidth,angle=0]{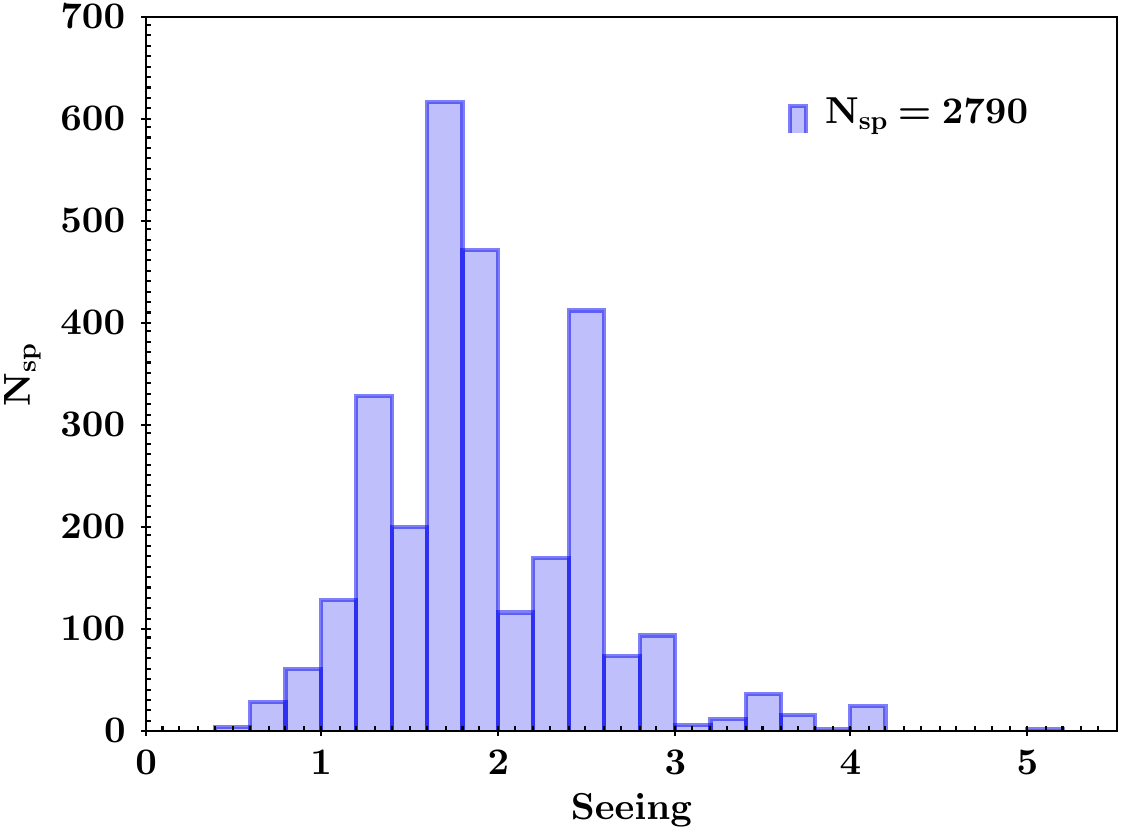}
\caption{Seeing distribution of library stars observations as reported in the GTC log files.}
\label{seeing}
\end{center}
\end{figure}

\section {Observations}
\label{observations}

We requested a total of 850\,h of open observing time; 700\,h were awarded in ten semesters, of which 487\,h were actually observed. The library observations were carried out within a filler-type programme. In line with the GTC guidelines, such programmes are scheduled when the seeing is worse than 1.2\,arcsec, under any nighttime conditions, including bright time and a range of sky qualities, with seeing being the main scheduling criterion. In this context, the library is especially suitable as a filler programme, since stellar observations can be performed under a wide range of atmospheric circumstances, which provides considerable flexibility for queue scheduling; in particular, the method adopted to derive the stellar atmospheric parameters  (Paper III) is primarily based on the relative properties of the spectral features rather than on the absolute flux scale, and thus photometric observing conditions are not required.

The seeing values are logged in the GTC files associated with MEGASTAR observations. In Fig.\,\ref{seeing}, we present the seeing distribution as a histogram, which shows a median of 1.9\,arcsec. Even for observations affected by relatively poor seeing, we retrieve the total flux by co-adding the spectra from 37 spaxels surrounding the fibre with the highest flux in the reconstructed IFU image. As a result, the spectral resolution is consistent across all targets, since the effective slit width (3.8 arcsec) is the same for every star, with the stop located at the microlens+fibre interface.

The observation strategy involves acquiring the \mbox{HR-R} and \mbox{HR-I} configurations within the same observation block (OB), thereby reducing overheads. In addition, for each OB the GTC provides observations of the spectrophotometric standard star to correct the instrumental response and perform flux calibration.  The reference spectra of these standard stars were obtained from  the European Southern Observatory (ESO) database\footnote{\url{https://www.eso.org/sci/observing/tools/standards/spectra/stanlis.html} }. The GTC observatory also provides
 halogen and ThNe lamp exposures, which are used for fibre tracing, flat-field correction, and wavelength calibration, respectively. The average exposure time per OB is 100\,s, as estimated with the MEGARA Exposure Time Calculator.  The faintest star observed within MEGASTAR has a V-band magnitude of 12.4. For the preparation of the GTC Phase 2, we rely on the MEGASTAR database and employ, among other tools, a visualisation interface to optimise target selection. We identify unobserved stars by applying filters on R and I magnitude ranges and/or on spectral type, as well as on any other stellar parameter.

As noted above, we found a deficit of metal-poor stars, cool giants, and late-type stars because these objects require long exposure times. Unfortunately, they were not chosen by the support astronomers even though we had prioritised and scheduled them during GTC Phase 2. Consequently, we have requested non-filler GTC open time to observe stars of specific spectral types to fill these gaps in the stellar parameter space. 

\section{Data reduction}
\label{datareduction}
\hyphenation{MegaraFiberFlatImage}
We processed the data using the publicly available MEGARA data reduction pipeline (DRP)\footnote{\url{https://github.com/guaix-ucm/megaradrp}}, developed by \citet{carandpas18, pasetal18, pasetal19} and  \citet{casetal18}.  The user executes a sequence of recipes to generate all required calibration frames. These recipes are  MegaraBiasImage, MegaraTraceMap, MegaraModelMap, MegaraArcCalibration, MegaraFiberFlatImage and MegaraLcbStdStar. Each recipe produces one or more output products, allowing the entire reduction process to be traced. Every routine relies on an input .yaml file that lists the images to be processed and defines the specific parameters required for that recipe. In addition, a global configuration file, control.yaml, stores all the information needed for data reduction, such as the polynomial order and the number of spectral lines to be included in the wavelength calibration.

The data reduction process begins with the generation of a  MasterBias using the MegaraBiasImage routine and the set of bias frames. All subsequent calibration files must be obtained using the same instrumental configuration. Cosmic-ray rejection is automatically applied by combining multiple exposures of the same target whenever available. The  MegaraTraceMap recipe uses halogen lamp exposures to identify the positions of the illuminated fibres on the detector and produce a DS9 region file for visual inspection. These regions can be overlaid on calibration and science frames to determine any offset between the reference traces and the actual fibre positions. Based on MegaraTraceMap output and halogen exposures, the  MegaraModelMap recipe performs the optimal extraction of the fibre spectra. Wavelength calibration is carried out with the MegaraArcCalibration recipe, which uses arc lamp exposures. The  MegaraFiberFlatImage recipe corrects for fibre-to-fibre transmission variations. The  MegaraLcbAcquisition recipe determines the position of the standard star on the LCB. The MegaraLcbStdStar routine derives the  master sensitivity curve by comparing the atmospheric-extinction-corrected 1D spectrum of the standard star with its tabulated flux-calibrated reference spectrum. This sensitivity function also accounts for the instrumental spectral response. Consequently, this calibration step remains necessary even under non-photometric observing conditions. An assessment of the relative flux calibration  accuracy was obtained from the dispersion root mean square (RMS) value) of independently derived sensitivity functions (master sensitivity curves). The internal consistency of 100 HR-R and 108  HR-I  derived sensitivity functions indicates a relative flux-calibration variation of approximately 3\% (see Appendix B).
 
The final data products consist of one Row-stacked-spectra (RSS) file per VPH, containing the relative-flux-calibrated (Jy) spectra for each individual fibre, already corrected for atmospheric extinction and the instrument response. To extract the 1D spectra, we use the reconstructed IFU  image produced by the Quick look application tool (QLA) \citep{gomezalv18}, and integrate over three concentric rings, each comprising 37 spaxels and centred on the brightest fibre, covering a total diameter of approximately 3.8\,arcsec. 

\label{datareduction}
\begin{figure}
\begin{center}
    \includegraphics[width=0.40\textwidth,angle=0]{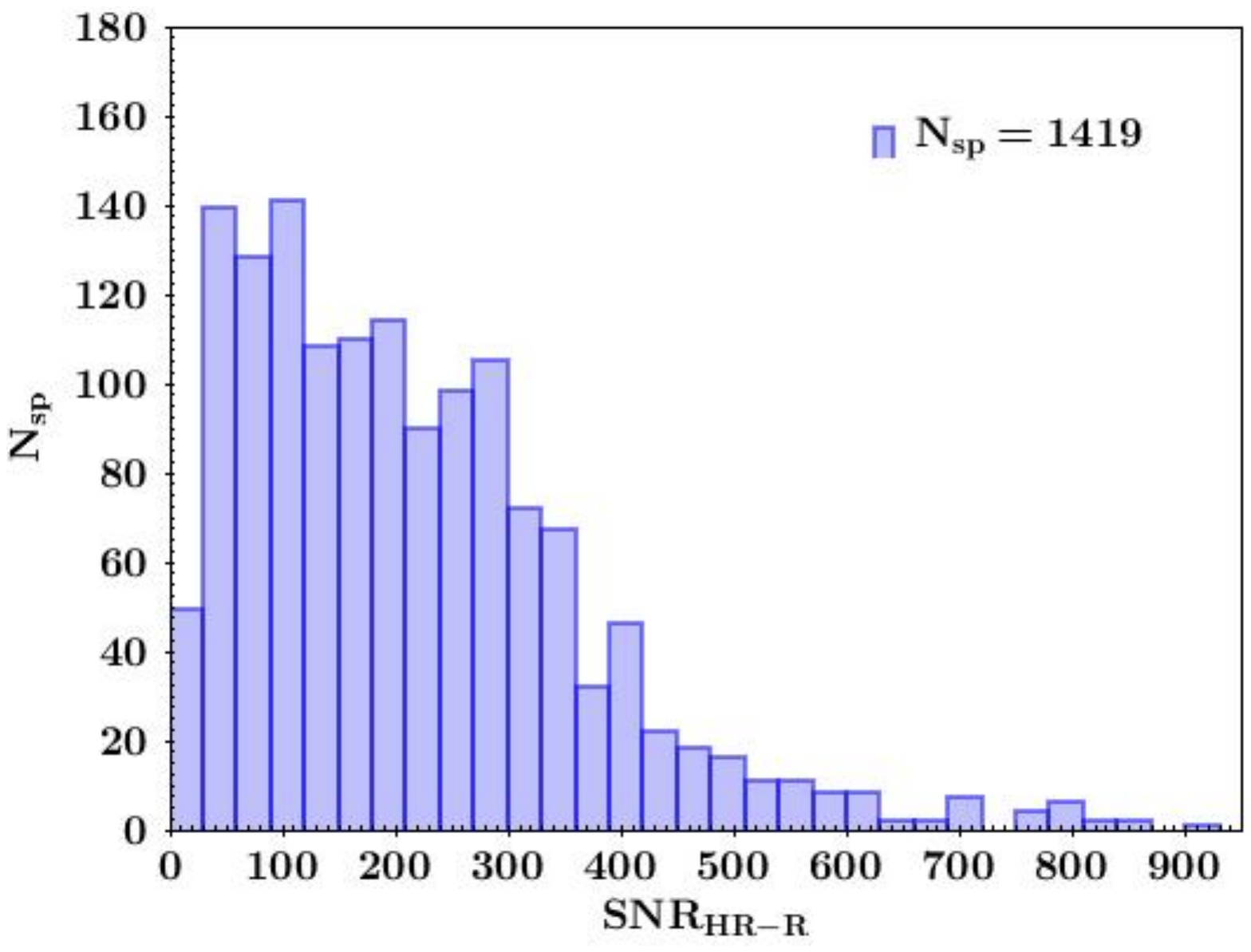}
\includegraphics[width=0.40\textwidth,angle=0]{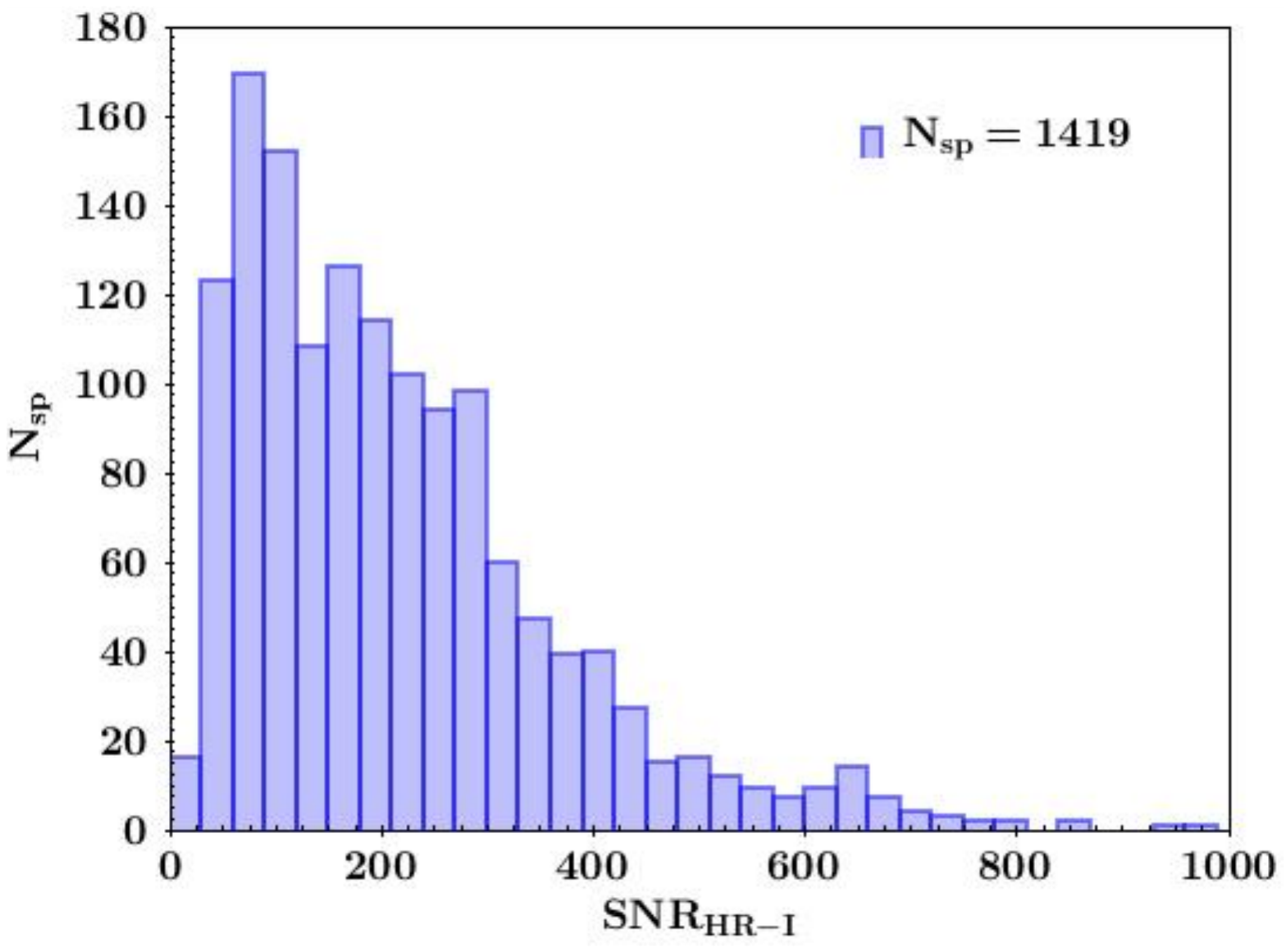}
\caption{Distribution of the S/N obtained from the observed spectra in the HR-R  (top) and HR-I (bottom) spectral configurations. The total number of spectra is indicated.}
\label{SNR}
\end{center}
\end{figure}

Fig. \ref{SNR} presents the continuum S/N distribution, computed following \citet[]{stoehr08}, for the HR-R and HR-I spectral setups. The mean S/N is approximately 215 in both cases, since the observations for the two configurations are obtained consecutively, with the integration time adjusted to the stellar colour. The broadening of the distributions directly results from the nature of the GTC filler-type programme, where observations are carried out whenever other programmes cannot proceed, primarily because of poor observing conditions. Nevertheless, filler observations can be scheduled simply to make full use of the night or to observe during twilight, and in some cases the conditions are actually quite favourable. In addition, each individual observation is carried out with the exposure time previously set in the GTC Phase 2 tool. In contrast, for other standard programmes the observing conditions are guaranteed, and the exposure time is then calculated to match those conditions. However, for MEGASTAR observations, the integration time has to be fixed in advance, independently of the actual observing conditions. At the same time, our library programme helps to optimise the use of GTC observing time, since without it this filler-type time would likely go unused. 

\begin{table}
\centering
\caption{Columns of the release\_summary\_DR2 file.}
\label{release-header}
\small
\begin{tabular}{lll} 
\hline
Column  & Description & \\
\hline
Name 	&	Star name (*) &\\			
RA	&	RA (2000.0) (hh:mm:ss.s) (*)& \\
DEC	&	DEC (2000.0) (dd:dd:ss.s) (*) &\\		
Sp Type  	&	 Spectral Type  (*)& \\				
U	&	Jonhson-Cousins U mag. (*) &\\		
B   &	Jonhson-Cousins B mag. (*)& \\			
V   &	Jonhson-Cousins V mag. (*)& \\			
R   &	Jonhson-Cousins R mag.(*)& \\			
I	&	Jonhson-Cousins I mag. (*) &\\			
J	&	Jonhson-Cousins J mag. (*) &\\			
Other name	&	Alternative name of the star& \\		
VPH &	 Grating used for observation &	\\
Teff	& Teff  (K) from literature &\\
Log\,g	&	Log\,g  from literature& \\
$[$Fe/H$]$	&	$[$Fe/H$]$ from literature &\\
Reference	& Original catalogue&\\
Other comments &   Comments  to star/observations &\\	
$\langle T_{\rm eff} \rangle$ &  Average of  valid models(**)&\\
$\Delta$ $\langle T_{\rm eff} \rangle$ &  Error in $\langle T_{\rm eff} \rangle$ &\\
$\langle \rm Log\,g \rangle$ &  Average of valid models(**)&\\
$\Delta$ $\langle \rm Log\,g \rangle$ & Error in $\langle \rm Log\,g \rangle$ &\\
$\langle {\rm [Fe/H]} \rangle$   &  Global average  valid models(**) &\\
$\Delta$ $\langle {\rm [Fe/H]} \rangle$  & Error in $\langle {\rm [Fe/H]} \rangle$ &\\
V$_{\rm rad}$  &  Radial velocity (km\,s$^{-1}$)(**)& \\
$\Delta$V$_{\rm rad}$  & Error in radial velocity &\\
<Reference> &    stellar parameters obtained from &\\
ASCII/FITS/ECSV file & ASCII/FITS/ECSV  formats&\\	
Obs. Period &	 GTC Observing semester& \\			
No. Exp  	&	 Number of exposures &\\				
Exp. Time   &	Individual exposure time (s) &\\
Seeing 	    &	Seeing values (arcsec) (***) &\\
Gaia(ID) DR3 &  Gaia DR3  Identifier &\\
Obs-GTC 	&	 GTC-OB identifier & \\
Rectified spectrum & Rectified spectrum availiability&\\
Products &  Data products availiability (DR1.1)&\\
\hline
(*) &Source: SIMBAD 4 Release 1.8&\\
(**) &See Paper\,III&\\
(***) &As reported in GTC log file&\\
 \end{tabular}
\end {table}

 The quality of sky subtraction in the MEGASTAR spectra is ensured by several factors. The eight sky mini-bundles, each consisting of 7 fibres, are located at an average distance of ~2 arcmin from the LCB, well beyond the wings of the stellar point spread function even for the brightest targets in the library. During data reduction, the individual sky spectra from each bundle are inspected and any bundle contaminated by an unrelated astrophysical source is excluded before computing the final sky reference spectrum. Given that all library targets are relatively bright \mbox{ (V $\leq$  12.4} mag), the sky background contribution to the total measured flux is negligible under all observing conditions, including bright lunar phases. Moreover, the HR-R and HR-I spectral windows do not contain any prominent sky emission lines that could compromise the stellar spectra. A careful visual inspection of all the reduced spectra confirms the absence of any systematic residuals attributable to imperfect sky subtraction. Further details on the calibration of the observations are provided in Appendix B.

\section{MEGASTAR database}
\label{database}

We implemented a MySQL database and an online tool to manage the stellar data and observed spectra released to the community as part of DR1 and DR1.1. More recently, we cross-matched the coordinates, proper motions, and magnitudes (U, B, V, R, I, J) with the SIMBAD~4 release 1.8. Furthermore, we searched for counterparts in Gaia DR3 by comparing positions between the MEGASTAR and Gaia DR3 catalogues. For this, we inspected a circular region with a radius of 2.4 arcsec, centred on the SIMBAD coordinates, corresponding to  about 1.2 times the MEGASTAR  effective slit width. 
A Gaia DR3 source ID was assigned when the mean offset in RA and DEC was less than or equal to 0.1 arcsec. Using this procedure, we identified Gaia DR3 counterparts for 1386 of the 1408 DR2 stars,  of which 43 have an exact match together with a second Gaia source within a radius of 2.4 arcsec. In these cases, the ID, coordinates, and G-magnitude of the secondary source are provided in the 'Other Comments' column of the DR2 summary table (release$\_$summary$\_$DR2). In the cross-matching process proper motions were taken into account.

The user can view the information for any individual star through the  Source menu, which contains all available stellar data, or perform a search by star name, spectral type, GTC phase-2, observed semester, VPH (HR-R or HR-I), Gaia DR3 ID, the originating library from which the star was selected, and the comments field to check whether the star has been observed. In addition, it is possible to search using ranges of coordinates, proper motions, magnitudes, and stellar parameters taken from the literature.

In the  Observations menu, users can search for individual stellar data or apply filters based on VPH, observing semester, and ranges of coordinates, number of exposures, exposure time, seeing, and spectrograph temperature (which is linked to the optimal focus configuration of the spectrograph), as well as on the comments and calibration comments fields. After choosing the spectral configuration, users can inspect the reduced and flux-calibrated spectrum and/or download it in ASCII, ECSV, and FITS formats. For DR1.1 stars, users may additionally download the rectified spectrum and a JPEG figure showing the best-fitting models from the theoretical grid, the residuals of those fits, and the stellar parameters derived from them, together with values available in the literature.

The  Products menu lists the products generated for DR1.1. Through the Find Products submenu, the user can search by stellar name or ranges in coordinates, T$_{\rm eff}$, log\,g, [Fe/H], S/N (HR-R and/or HR-I), as well as by the stellar indices defined in Paper\,III. 
 When a filter is applied to any of these fields, the application generates a results table available in PDF, ASCII, and MsExcel formats. In addition, via the Product Tables submenu, the user can download a complete table containing all stars for which stellar parameters and spectral indices have been derived. The  Download menu enables the user to obtain MEGASTAR DR1, DR1.1, and DR2.

\section{MEGASTAR data release 2}
\label{release}

The DR2 is provided as a zip package containing: (i) the instructions; (ii) a file (release$\_$summary$\_$DR2) with detailed information on the spectra delivered for each star; (iii) a zip archive with the spectra; and (iv) a Java application to visualise the observations. After extracting the main zip file, four subfolders are created, each holding the 2838 spectra in a distinct format (ASCII, ECSV, FITS, and JPG). Lists of the spectrum filenames for each of the four subfolders are also included.
   
In Table~\ref{release-header}, the description of the columns  of the file release$\_$summary$\_$DR2 is provided. The  catalogue of DR2 stars is presented in  Table \ref{starslist} in Appendix A. Details on how to access the database, from which MEGASTAR DR2 can be downloaded, are provided in section \ref {availiability}.
 
\section {Concluding remarks}
\label {summary}
We presented DR2 of MEGASTAR, a stellar spectral library constructed using the MEGARA IFU. This release provides 2838 spectra for 1408 stars, all obtained with the high-resolution mode, achieving a spectral resolution of R(FWHM)\,$\sim$\,20000. The HR-R and HR-I setups cover wavelength intervals of 
6420\,–\,6790\,\AA\, and 8370\,–\,8885\,\AA, respectively. The observations were carried out at the GTC in filler mode over the course of ten semesters. Data were uniformly processed using the MEGARA data reduction pipeline. Because the programme was executed in filler mode, the median seeing was 1.9 arcsec. Even under these relatively poor seeing conditions, we recovered most of the flux by extracting each spectrum from 37 spaxels, centred on the spaxel with the highest intensity. The average S/N of the spectra is about 215 in both spectral setups.  The spectra are provided in Jy following  flux calibration relative to the standard-star observations obtained  under non-photometric conditions; since one of our primary goals is to determine stellar parameters in a homogeneous manner for all stars in the library, a reliable relative flux calibration is sufficient as the analysis is based on continuum-rectified spectra and on the comparison of spectral features with theoretical models (see Paper III) rather than on the absolute flux scale.

We have developed a dedicated database, a web-based application, and a graphical interface that enable users to access and visualise the spectra. Through these tools, the MEGASTAR spectra and related products are available to the community. By logging in users can obtain data for 1408 stars, either by retrieving or inspecting the individual spectra obtained with the two spectral configurations, HR-R and HR-I, and/or by downloading the complete data release. Furthermore, to facilitate the effective exploitation of the released data and derived products, we have upgraded and refined the MEGASTAR database and its web-based tools.

We emphasise that one of the major strengths of our library is that it was constructed using an IFU. As a result, the spectral resolution remains constant since the slit width is identical for every star, given that we systematically extracted the spectra from 37 spaxels in the reconstructed  IFU image. Since the first MEGASTAR release, which included observations of 414 stars, an additional 994 stars have been observed, significantly enhancing our coverage of the stellar parameter space. Consequently, MEGASTAR meets the criteria for a modern empirical library as defined by \citet{trager12}: it is well calibrated, free from slit effects, includes observations of a large number of stars, and delivers sufficiently high spectral resolution to model both individual stellar clusters and galaxies. We expect MEGASTAR to be valuable for a variety of projects. In particular, our primary scientific objective is to generate MEGARA SSP blocks using MEGASTAR, in combination with the evolutionary synthesis models of {\sc HR-PyPopStar} \citep{Millan2021, Millan2025}.

\section{Data availability}
\label{availiability}
The table release$\_$summary$\_$DR2, whose columns are described in Table~\ref{release-header}, is available only in electronic form at the CDS via anonymous ftp to cdsarc.u-strasbg.fr (130.79.128.5) or via http://cdsweb.u-strasbg.fr/cgi-bin/qcat?J/A+A/.

The MEGASTAR DR2 spectra and products are available in the library database,\footnote {\url {https://www.fractal-es.com/megaragtc-stellarlibrary/}} username: {public};  password: {Q50ybAZm}.

\begin{acknowledgements}
The authors are grateful to the referee for the careful reading of the manuscript and for constructive comments and suggestions, which have significantly improved the quality and clarity of the paper.
This work has been partially funded by FRACTAL, INAOE and CIEMAT. This work is based on observations made with the Gran Telescopio Canarias (GTC), installed at the Spanish Observatorio del Roque de los Muchachos, in the island of La Palma. This work is based on data obtained with the MEGARA instrument, funded by the European Regional Development Fund (ERDF), through Programa Operativo Canarias FEDER 2014-2020. The authors thank Dr. Antonio Cabrera and Dr. Daniel Reverte, GTC Operations Group staff, for the support provided during the preparation and execution of the observations at the GTC. This research made use of Astropy, a community-developed core Python package for Astronomy. This work has used data from the European Space Agency (ESA) Gaia mission
 (\url{https://www.cosmos.esa.int/gaia}), processed by the  Gaia
Data Processing and Analysis Consortium (DPAC,
\url{https://www.cosmos.esa.int/web/gaia/dpac/consortium}). Funding for the DPAC
has been provided by national institutions, in particular the institutions
participating in the  Multilateral Agreement.
E. Carrasco, M. L. Garc\'{i}a-Vargas and M. Moll\'{a} acknowledge funding from grant PID2022-136598NB-C33 funded by MCIN/AEI/10.13039/501100011033 and by "ERDF A way of making Europe". A. Gil de Paz and N. Cardiel acknowledge financial support from the Spanish MICIN under grants PID2022-138621NB-I00 and PID2021-123417OB-I00. S. Rodr\'{i}guez-Berlanas acknowledges funding from the Spanish Ministry of Science and Innovation (MICINN) through the Spanish State Research Agency through
grants PID2021-122397NB-C21 and the Severo Ochoa Programme 2020-2024 (CEX2019-000920-S),  and to NextGeneration EU/PRTR and MIU (UNI/551/2021) through the Margarita Salas-ULL grant.
\end{acknowledgements}

\onecolumn
\begin{appendix}
\section{DR2 catalogue}
\begin{longtable}  {llllllll}
\endfirsthead
\caption{continued.}\\
 \hline \hline
 &&&&&&&\\
\endhead
\hline
\endfoot
\caption{\label{starslist} List of the stars in MEGASTAR DR2.}\\
\hline \hline
&&&&&&&\\
** KIR 4	&	BD+463471	&	HD014633	&	HD027819	&	HD036395	&	HD041117	&	HD044030	&	HD048682B	\\
BD+012916	&	BD+473458	&	HD014947	&	HD027887	&	HD036512	&	HD041142	&	HD044033	&	HD048737	\\
BD+053080	&	BD+511696	&	HD015318	&	HD027962	&	HD036667	&	HD041307	&	HD044109	&	HD048781	\\
BD+062986	&	BD+541399	&	HD015558	&	HD028005	&	HD036673	&	HD041330	&	HD044195	&	HD048948	\\
BD+083095	&	BD+800245	&	HD015570	&	HD028100	&	HD036861	&	HD041357	&	HD044274	&	HD049161	\\
BD+092190	&	BD-032525	&	HD015629	&	HD028307	&	HD036960	&	HD041361	&	HD044333	&	HD049178	\\
BD+092860	&	BD-122669	&	HD016429	&	HD028343	&	HD037124	&	HD041433	&	HD044478	&	HD049330	\\
BD+112998	&	BD-185550	&	HD016523	&	HD028527	&	HD037202	&	HD041460	&	HD044515	&	HD049367	\\
BD+130013	&	G115-58	&	HD016581	&	HD028546	&	HD037216	&	HD041501	&	HD044537	&	HD049368	\\
BD+151305	&	G156-031	&	HD017081	&	HD028946	&	HD037269	&	HD041593	&	HD044602	&	HD049409	\\
BD+172473	&	G169-28	&	HD017145	&	HD028978	&	HD037272	&	HD041636	&	HD044614	&	HD049520	\\
BD+182890	&	G171-010	&	HD017506	&	HD029150	&	HD037394	&	HD041640	&	HD044638	&	HD049601	\\
BD+182976	&	G192-43	&	HD017638	&	HD029574	&	HD037526	&	HD041692	&	HD044770	&	HD049732	\\
BD+191730	&	G197-45	&	HD018078	&	HD029645	&	HD037536	&	HD041712	&	HD044867	&	HD049739	\\
BD+195116B	&	G202-65	&	HD018144	&	HD029697	&	HD037742	&	HD041808	&	HD044889	&	HD049932	\\
BD+203603	&	G234-28	&	HD018296	&	HD030034	&	HD037788	&	HD042035	&	HD044947	&	HD050277	\\
BD+233130	&	G63-26	&	HD018409	&	HD030210	&	HD037958	&	HD042089	&	HD044985	&	HD050420	\\
BD+241676	&	GJ 1111	&	HD018769	&	HD030504	&	HD038145	&	HD042092	&	HD045152	&	HD050522	\\
BD+251981	&	GJ 1245A	&	HD019308	&	HD030614	&	HD038230	&	HD042111	&	HD045205	&	HD050554	\\
BD+252436	&	GJ 1245B	&	HD019510	&	HD030649	&	HD038237	&	HD042143	&	HD045317	&	HD050692	\\
BD+252479	&	HD000108	&	HD020041	&	HD030676	&	HD038247	&	HD042182	&	HD045321	&	HD050696	\\
BD+252497	&	HD000358	&	HD020084	&	HD030834	&	HD038270	&	HD042217	&	HD045355	&	HD050794	\\
BD+262276	&	HD000560	&	HD020468	&	HD030959	&	HD038309	&	HD042250	&	HD045391	&	HD051219	\\
BD+262606	&	HD000886	&	HD020512	&	HD031219	&	HD038528	&	HD042256	&	HD045410	&	HD051309	\\
BD+272057	&	HD001400	&	HD021742	&	HD031236	&	HD038529	&	HD042299	&	HD045412	&	HD051419	\\
BD+272096	&	HD001918	&	HD022484	&	HD031293	&	HD038545	&	HD042353	&	HD045431	&	HD051440	\\
BD+272126	&	HD002126	&	HD023183	&	HD031295	&	HD038650	&	HD042454	&	HD045507	&	HD051530	\\
BD+282079	&	HD002628	&	HD023194	&	HD031374	&	HD038656	&	HD042474	&	HD045514	&	HD052005	\\
BD+292091	&	HD002796	&	HD023607	&	HD031412	&	HD038751	&	HD042475	&	HD045759	&	HD052711	\\
BD+292231	&	HD002857	&	HD023862	&	HD031966	&	HD038798	&	HD042543	&	HD045829	&	HD052973	\\
BD+292290	&	HD002952	&	HD023924	&	HD031996	&	HD038856	&	HD042548	&	HD045910	&	HD053927	\\
BD+292294	&	HD003360	&	HD024341	&	HD032259	&	HD038899	&	HD042597	&	HD046122	&	HD053929	\\
BD+302431	&	HD003369	&	HD024451	&	HD032537	&	HD039003	&	HD042618	&	HD046199	&	HD054217	\\
BD+302611	&	HD003574	&	HD024496	&	HD032655	&	HD039082	&	HD042773	&	HD046223	&	HD054322	\\
BD+302931	&	HD003628	&	HD024534	&	HD032736	&	HD039118	&	HD042787	&	HD046317	&	HD054357	\\
BD+312356	&	HD003644	&	HD024912	&	HD032923	&	HD039225	&	HD042807	&	HD046380	&	HD054371	\\
BD+312360	&	HD003765	&	HD025173	&	HD033299	&	HD039587	&	HD042983	&	HD046454	&	HD054717	\\
BD+321561	&	HD003817	&	HD025291	&	HD033632	&	HD039681	&	HD042995	&	HD046480	&	HD054719	\\
BD+342371	&	HD004004	&	HD025329	&	HD033904	&	HD039773	&	HD043021	&	HD046588	&	HD055052	\\
BD+342476	&	HD004539	&	HD025532	&	HD033959	&	HD039801	&	HD043039	&	HD046687	&	HD055280	\\
BD+351484	&	HD004778	&	HD025621	&	HD034078	&	HD039866	&	HD043042	&	HD046703	&	HD055383	\\
BD+362165	&	HD004963	&	HD025825	&	HD034255	&	HD039882	&	HD043083	&	HD046784	&	HD055575	\\
BD+362219	&	HD005294	&	HD025893	&	HD034334	&	HD039949	&	HD043153	&	HD047072	&	HD055606	\\
BD+362303	&	HD005516	&	HD026322	&	HD034578	&	HD039970	&	HD043188	&	HD047105	&	HD056359	\\
BD+363168	&	HD006327	&	HD026462	&	HD034797	&	HD040111	&	HD043247	&	HD047127	&	HD056925	\\
BD+371458	&	HD006815	&	HD026574	&	HD034816	&	HD040183	&	HD043264	&	HD047129	&	HD056986	\\
BD+371665	&	HD006961	&	HD026736	&	HD034878	&	HD040239	&	HD043285	&	HD047309	&	HD057264	\\
BD+372312	&	HD007374	&	HD026756	&	HD035468	&	HD040259	&	HD043286	&	HD047731	&	HD057275	\\
BD+372319	&	HD008574	&	HD026923	&	HD035497	&	HD040335	&	HD043338	&	HD047839	&	HD057669	\\
BD+381670	&	HD009974	&	HD027126	&	HD035601	&	HD040372	&	HD043358	&	HD047914	&	HD057727	\\
BD+392554	&	HD009996	&	HD027282	&	HD035620	&	HD040446	&	HD043378	&	HD048099	&	HD058072	\\
BD+394926	&	HD010975	&	HD027295	&	HD035770	&	HD040460	&	HD043380	&	HD048144	&	HD058207	\\
BD+412695	&	HD011749	&	HD027371	&	HD035961	&	HD040573	&	HD043384	&	HD048279	&	HD058343	\\
BD+413306	&	HD013267	&	HD027382	&	HD036066	&	HD040650	&	HD043523	&	HD048329	&	HD058551	\\
BD+422667	&	HD013268	&	HD027524	&	HD036130	&	HD040801	&	HD043526	&	HD048433	&	HD058595	\\
BD+451668	&	HD013825	&	HD027561	&	HD036165	&	HD040964	&	HD043587	&	HD048565	&	HD058781	\\
BD+452247	&	HD014191	&	HD027685	&	HD036215	&	HD040996	&	HD043856	&	HD048568	&	HD058855	\\
BD+461635	&	HD014489	&	HD027697	&	HD036389	&	HD041079	&	HD043947	&	HD048682	&	HD058923	\\
HD058946	&	HD071515	&	HD078362	&	HD089125	&	HD098839	&	HD108678	&	HD116842	&	HD128987	\\
HD059058	&	HD071557	&	HD078418	&	HD089251	&	HD099028	&	HD108915	&	HD117176	&	HD129132	\\
HD059374	&	HD071881	&	HD078479	&	HD089269	&	HD099329	&	HD108945	&	HD117243	&	HD129174	\\
HD059473	&	HD071952	&	HD078712	&	HD089307	&	HD099505	&	HD108954	&	HD117858	&	HD129312	\\
HD059881	&	HD072097	&	HD079028	&	HD089389	&	HD099648	&	HD108956	&	HD117876	&	HD129336	\\
HD060111	&	HD072184	&	HD079158	&	HD089449	&	HD099747	&	HD108985	&	HD118022	&	HD129814	\\
HD060178J	&	HD072324	&	HD079210	&	HD089484	&	HD100006	&	HD108986	&	HD118096	&	HD129956	\\
HD060179	&	HD072505	&	HD079283	&	HD089744	&	HD100030	&	HD109303	&	HD118244	&	HD129972	\\
HD060319	&	HD072660	&	HD079452	&	HD089758	&	HD100284	&	HD109307	&	HD118266	&	HD130025	\\
HD060408	&	HD072722	&	HD079469	&	HD089822	&	HD100446	&	HD109317	&	HD118905	&	HD130087	\\
HD060501	&	HD072905	&	HD079555	&	HD089995	&	HD100470	&	HD109345	&	HD119228	&	HD130109	\\
HD060522	&	HD072945	&	HD079633	&	HD090250	&	HD100563	&	HD109358	&	HD119288	&	HD130705	\\
HD061295	&	HD072946	&	HD079765	&	HD090277	&	HD100696	&	HD109995	&	HD119291	&	HD130817	\\
HD061603	&	HD072968	&	HD080081	&	HD090343	&	HD101013	&	HD110184	&	HD119516	&	HD130948	\\
HD061606	&	HD073226	&	HD080218	&	HD090361	&	HD101107	&	HD110679	&	HD119550	&	HD131042	\\
HD061913	&	HD073344	&	HD080390	&	HD090507	&	HD101177	&	HD110833	&	HD119802	&	HD131111	\\
HD062301	&	HD073394	&	HD080536	&	HD090508	&	HD101177B	&	HD110885	&	HD119850	&	HD131156A	\\
HD062345	&	HD073593	&	HD080715	&	HD090537	&	HD101227	&	HD110897	&	HD120066	&	HD131156B	\\
HD062346	&	HD073665	&	HD081029	&	HD090839	&	HD101484	&	HD110930	&	HD120136	&	HD131507	\\
HD062437	&	HD073667	&	HD081146	&	HD090862	&	HD101501	&	HD111028	&	HD120164	&	HD132254	\\
HD062613	&	HD073668	&	HD081192	&	HD091190	&	HD101606	&	HD111154	&	HD120315	&	HD132756	\\
HD062721	&	HD073710	&	HD082106	&	HD091204	&	HD101676	&	HD111398	&	HD120348	&	HD132933	\\
HD063302	&	HD074000	&	HD082210	&	HD091316	&	HD101690	&	HD111515	&	HD120933	&	HD133124	\\
HD063333	&	HD074011	&	HD082395	&	HD091347	&	HD102224	&	HD111591	&	HD121370	&	HD133165	\\
HD063352	&	HD074156	&	HD082635	&	HD091752	&	HD102328	&	HD111812	&	HD121560	&	HD133208	\\
HD063410	&	HD074280	&	HD082741	&	HD092125	&	HD102780	&	HD112030	&	HD121934	&	HD134044	\\
HD063433	&	HD074377	&	HD083425	&	HD092127	&	HD102870	&	HD112127	&	HD122120	&	HD134063	\\
HD063778	&	HD074442	&	HD083632	&	HD092769	&	HD103095	&	HD112300	&	HD122408	&	HD134083	\\
HD063791	&	HD074462	&	HD083787	&	HD093329	&	HD103578	&	HD112413	&	HD122563	&	HD134113	\\
HD064090	&	HD074721	&	HD083805	&	HD093430	&	HD103877	&	HD112735	&	HD122742	&	HD134169	\\
HD064332	&	HD074739	&	HD084347	&	HD093487	&	HD104556	&	HD113002	&	HD122796	&	HD134190	\\
HD064412	&	HD075302	&	HD084441	&	HD093765	&	HD104783	&	HD113022	&	HD123299	&	HD134474	\\
HD064488	&	HD075318	&	HD084453	&	HD094028	&	HD104979	&	HD113095	&	HD123657	&	HD135101	\\
HD064606	&	HD075332	&	HD084737	&	HD094264	&	HD104985	&	HD113226	&	HD123821	&	HD135148	\\
HD064630	&	HD075333	&	HD084748	&	HD094587	&	HD105087	&	HD113285	&	HD123977	&	HD135485	\\
HD065123	&	HD075732	&	HD084937	&	HD094600	&	HD105262	&	HD113319	&	HD124186	&	HD135599	\\
HD065372	&	HD075782	&	HD084971	&	HD094601	&	HD105546	&	HD113436	&	HD124320	&	HD135722	\\
HD065583	&	HD075933	&	HD085235	&	HD094669	&	HD105740	&	HD113515	&	HD124570	&	HD136202	\\
HD065714	&	HD076219	&	HD086133	&	HD094672	&	HD105755	&	HD113847	&	HD125451	&	HD136512	\\
HD065804	&	HD076221	&	HD086322	&	HD094705	&	HD105944	&	HD113848	&	HD125560	&	HD136514	\\
HD065900	&	HD076291	&	HD086560	&	HD094835	&	HD106038	&	HD113996	&	HD126031	&	HD136923	\\
HD066011	&	HD076292	&	HD086728	&	HD095128	&	HD106156	&	HD114092	&	HD126053	&	HD137071	\\
HD066090	&	HD076351	&	HD086873	&	HD095241	&	HD106210	&	HD114174	&	HD126141	&	HD137391	\\
HD066444	&	HD076508	&	HD086986	&	HD095418	&	HD106252	&	HD114256	&	HD126246	&	HD137471	\\
HD066573	&	HD076572	&	HD087140	&	HD095735	&	HD106365	&	HD114357	&	HD126271	&	HD137510	\\
HD067228	&	HD076752	&	HD087141	&	HD096094	&	HD106591	&	HD114606	&	HD126327	&	HD137704	\\
HD067390	&	HD076780	&	HD087344	&	HD096360	&	HD106714	&	HD114710	&	HD126511	&	HD137909	\\
HD067767	&	HD076813	&	HD087646	&	HD096436	&	HD107213	&	HD114762	&	HD126512	&	HD138279	\\
HD068017	&	HD076943	&	HD087696	&	HD096833	&	HD107276	&	HD115004	&	HD126660	&	HD138290	\\
HD068638	&	HD076944	&	HD087737	&	HD097560	&	HD107328	&	HD115136	&	HD126661	&	HD138481	\\
HD068988	&	HD077093	&	HD087822	&	HD097603	&	HD107383	&	HD115268	&	HD126778	&	HD138573	\\
HD069897	&	HD077350	&	HD087901	&	HD097633	&	HD107513	&	HD115274	&	HD127243	&	HD138749	\\
HD070272	&	HD077601	&	HD088072	&	HD097711	&	HD107582	&	HD115383	&	HD127334	&	HD138764	\\
HD070298	&	HD077729	&	HD088230	&	HD097778	&	HD107700	&	HD115404	&	HD127506	&	HD139195	\\
HD070458	&	HD077818	&	HD088355	&	HD097855	&	HD107705	&	HD115444	&	HD127665	&	HD139323	\\
HD070757	&	HD077912	&	HD088446	&	HD097907	&	HD107752	&	HD115539	&	HD128000	&	HD139324	\\
HD071030	&	HD078175	&	HD088609	&	HD097916	&	HD107950	&	HD115604	&	HD128165	&	HD139341	\\
HD071148	&	HD078209	&	HD088725	&	HD098262	&	HD108076	&	HD116316	&	HD128167	&	HD139457	\\
HD071155	&	HD078234	&	HD088737	&	HD098468	&	HD108177	&	HD116442	&	HD128661	&	HD139641	\\
HD071262	&	HD078249	&	HD088986	&	HD098744	&	HD108225	&	HD116443	&	HD128750	&	HD139798	\\
HD071310	&	HD078316	&	HD089010	&	HD098824	&	HD108381	&	HD116515	&	HD128801	&	HD140027	\\
HD140160	&	HD152792	&	HD163990	&	HD173780	&	HD183915	&	HD194093	&	HD210889	&	HD224926	\\	
HD140232	&	HD152815	&	HD163993	&	HD173799	&	HD184385	&	HD194244	&	HD211075	&	HD225160	\\	
HD140233	&	HD152830	&	HD164136	&	HD173819	&	HD184499	&	HD194453	&	HD211472	&	HD232078	\\	
HD140573	&	HD153286	&	HD164349	&	HD173936	&	HD184786	&	HD195198	&	HD212061	&	HD232979	\\	
HD141003	&	HD153808	&	HD164353	&	HD174069	&	HD184875	&	HD195324	&	HD212076	&	HD233345	\\	
HD141004	&	HD153882	&	HD164432	&	HD174512	&	HD184927	&	HD195506	&	HD212442	&	HD233511	\\	
HD141272	&	HD154278	&	HD164557	&	HD174567	&	HD185351	&	HD195592	&	HD212454	&	HD233666	\\	
HD141680	&	HD154345	&	HD164922	&	HD174701	&	HD185657	&	HD195729	&	HD212496	&	HD233891	\\	
HD141714	&	HD154431	&	HD165029	&	HD174884	&	HD185859	&	HD196426	&	HD212571	&	HD234677	\\	
HD141795	&	HD154660	&	HD165195	&	HD174912	&	HD185936	&	HD196610	&	HD212593	&	HD237846	\\	
HD142091	&	HD154733	&	HD165358	&	HD174959	&	HD186307	&	HD196662	&	HD213420	&	HD237903	\\	
HD142093	&	HD154931	&	HD165645	&	HD175037	&	HD186377	&	HD196725	&	HD213470	&	HD241253	\\	
HD142373	&	HD155358	&	HD165670	&	HD175100	&	HD186379	&	HD196777	&	HD214080	&	HD250792	\\	
HD142575	&	HD155514	&	HD165887	&	HD175156	&	HD186412	&	HD197345	&	HD214167	&	HD251611	\\	
HD142860	&	HD155763	&	HD166014	&	HD175306	&	HD186486	&	HD197572	&	HD214168	&	HD277559	\\	
HD142908	&	HD156164	&	HD166046	&	HD175376	&	HD186619	&	HD197812	&	HD214680	&	HD281679	\\	
HD142926	&	HD156283	&	HD166207	&	HD175426	&	HD186675	&	HD197912	&	HD214994	&	HD284248	\\	
HD142980	&	HD156729	&	HD166208	&	HD175535	&	HD186980	&	HD197989	&	HD215373	&	HD338529	\\	
HD143107	&	HD157198	&	HD166229	&	HD175588	&	HD187013	&	HD198183	&	HD215512	&	HR5511	\\	
HD143213	&	HD157214	&	HD166620	&	HD175640	&	HD187111	&	HD198345	&	HD215665	&	LHS10	\\	
HD143393	&	HD157373	&	HD166734	&	HD175786	&	HD187123	&	HD198478	&	HD215704	&	NGC 2682 156	\\	
HD143666	&	HD157466	&	HD166958	&	HD175865	&	HD187216	&	HD199191	&	HD216143	&	NGC 2682 55	\\	
HD143761	&	HD157857	&	HD167006	&	HD176301	&	HD187282	&	HD199478	&	HD216174	&	NGC2420-140	\\	
HD143807	&	HD157910	&	HD167042	&	HD176318	&	HD187459	&	HD199579	&	HD216672	&	NGP29129	\\	
HD144206	&	HD158148	&	HD167392	&	HD176437	&	HD187811	&	HD199799	&	HD216831	&	Ross-889	\\	
HD144284	&	HD158261	&	HD167588	&	HD176582	&	HD187879	&	HD199870	&	HD216916	&	TYC 3015-1951-1	\\	
HD144287	&	HD158414	&	HD167771	&	HD176630	&	HD187921	&	HD200527	&	HD217086	&	V* BN Vul	\\	
HD144579	&	HD158899	&	HD168009	&	HD176819	&	HD187929	&	HD200580	&	HD217382	&	V* GK Com	\\	
HD144872	&	HD159062	&	HD168322	&	HD177196	&	HD187983	&	HD200905	&	HD217476	&	V* IW Com	\\	
HD145148	&	HD159139	&	HD168608	&	HD177282	&	HD188001	&	HD201091	&	HD217833	&		\\	
HD145328	&	HD159181	&	HD168720	&	HD177880	&	HD188056	&	HD201626	&	HD217891	&		\\	
HD145389	&	HD159222	&	HD168775	&	HD178187	&	HD188209	&	HD201889	&	HD217906	&		\\	
HD145502	&	HD159332	&	HD168797	&	HD178287	&	HD188260	&	HD202109	&	HD218031	&		\\	
HD145675	&	HD159482	&	HD169027	&	HD178329	&	HD188326	&	HD202124	&	HD218045	&		\\	
HD145729	&	HD159909	&	HD169305	&	HD178359	&	HD188485	&	HD202573	&	HD218059	&		\\	
HD145976	&	HD160054	&	HD169414	&	HD178717	&	HD188650	&	HD202671	&	HD218329	&		\\	
HD147394	&	HD160365	&	HD169822	&	HD179104	&	HD188947	&	HD203344	&	HD218376	&		\\	
HD147550	&	HD160693	&	HD169931	&	HD179315	&	HD189087	&	HD204543	&	HD218915	&		\\	
HD147677	&	HD160762	&	HD170009	&	HD179406	&	HD189479	&	HD204642	&	HD218935	&		\\	
HD147923	&	HD160910	&	HD170737	&	HD179761	&	HD190073	&	HD204754	&	HD219734	&		\\	
HD148112	&	HD161056	&	HD170739	&	HD180006	&	HD190229	&	HD204771	&	HD219945	&		\\	
HD148184	&	HD161074	&	HD170756	&	HD180029	&	HD190603	&	HD204867	&	HD219962	&		\\	
HD148816	&	HD161149	&	HD170780	&	HD180124	&	HD190864	&	HD205021	&	HD219978	&		\\	
HD148856	&	HD161239	&	HD170783	&	HD180163	&	HD191046	&	HD205331	&	HD220182	&		\\	
HD148897	&	HD161573	&	HD170973	&	HD180554	&	HD191615	&	HD205512	&	HD220575	&		\\	
HD149009	&	HD161603	&	HD171301	&	HD180610	&	HD191692	&	HD206165	&	HD220787	&		\\	
HD149121	&	HD161677	&	HD171406	&	HD180928	&	HD191765	&	HD206267	&	HD220825	&		\\	
HD149161	&	HD161695	&	HD171995	&	HD181096	&	HD192103	&	HD206374	&	HD220933	&		\\	
HD149382	&	HD161796	&	HD172380	&	HD181276	&	HD192163	&	HD206778	&	HD221170	&		\\	
HD149419	&	HD161797A	&	HD172488	&	HD181440	&	HD192281	&	HD206859	&	HD221345	&		\\	
HD149750	&	HD161817	&	HD172816	&	HD181470	&	HD192577	&	HD207076	&	HD221585	&		\\	
HD149757	&	HD161941	&	HD172958	&	HD181475	&	HD192639	&	HD207134	&	HD221830	&		\\	
HD150012	&	HD162028	&	HD173034	&	HD181655	&	HD192641	&	HD207330	&	HD222107	&		\\	
HD150449	&	HD162211	&	HD173158	&	HD181751	&	HD192907	&	HD207673	&	HD223047	&		\\	
HD150680	&	HD162555	&	HD173329	&	HD182272	&	HD192909	&	HD208501	&	HD223165	&		\\	
HD150997	&	HD162570	&	HD173399	&	HD182293	&	HD192944	&	HD209459	&	HD223385	&		\\	
HD151044	&	HD162753	&	HD173524	&	HD182488	&	HD193077	&	HD209975	&	HD223869	&		\\	
HD151203	&	HD162954	&	HD173650	&	HD182568	&	HD193322	&	HD210424	&	HD224458	&		\\	
HD151288	&	HD163506	&	HD173667	&	HD182736	&	HD193432	&	HD210745	&	HD224544	&		\\	
HD151877	&	HD163588	&	HD173673	&	HD182761	&	HD193576	&	HD210809	&	HD224559	&		\\	
HD152614	&	HD163641	&	HD173701	&	HD183144	&	HD193793	&	HD210839	&	HD224801	&		\\	
\end{longtable}

\twocolumn
\section{Calibrations}
\subsection{Flux calibration accuracy}
An assessment of the relative flux calibration  accuracy was obtained from the dispersion of independently derived sensitivity functions (master sensitivity curves), generated by the MEGARA data reduction pipeline, for the HR-R and HR-I configurations. The  MegaraLcbStdStar routine derives these curves by comparing the atmospheric-extinction-corrected 1D spectrum of the standard star with its tabulated flux-calibrated reference spectrum. This sensitivity function also accounts for the instrumental spectral response. This calibration step is indispensable even when the observations have been carried out under non-photometric conditions.  
\begin{figure}[!ht]

Before statistical analysis, each sensitivity function was normalised by its integrated flux to compare only the relative shape of the calibration curves and to remove differences in absolute scaling, as shown in Fig.\ref{sensitivity}. For the HR-R setup, the RMS computed over the full wavelength range from a sample of 100 sensitivity curves is 3.46\%, while for the \HRI\ setup the corresponding RMS, derived from 108 sensitivity curves, is 3.27\%. 

\begin{center}
    \includegraphics[width=0.88\columnwidth,angle=0]{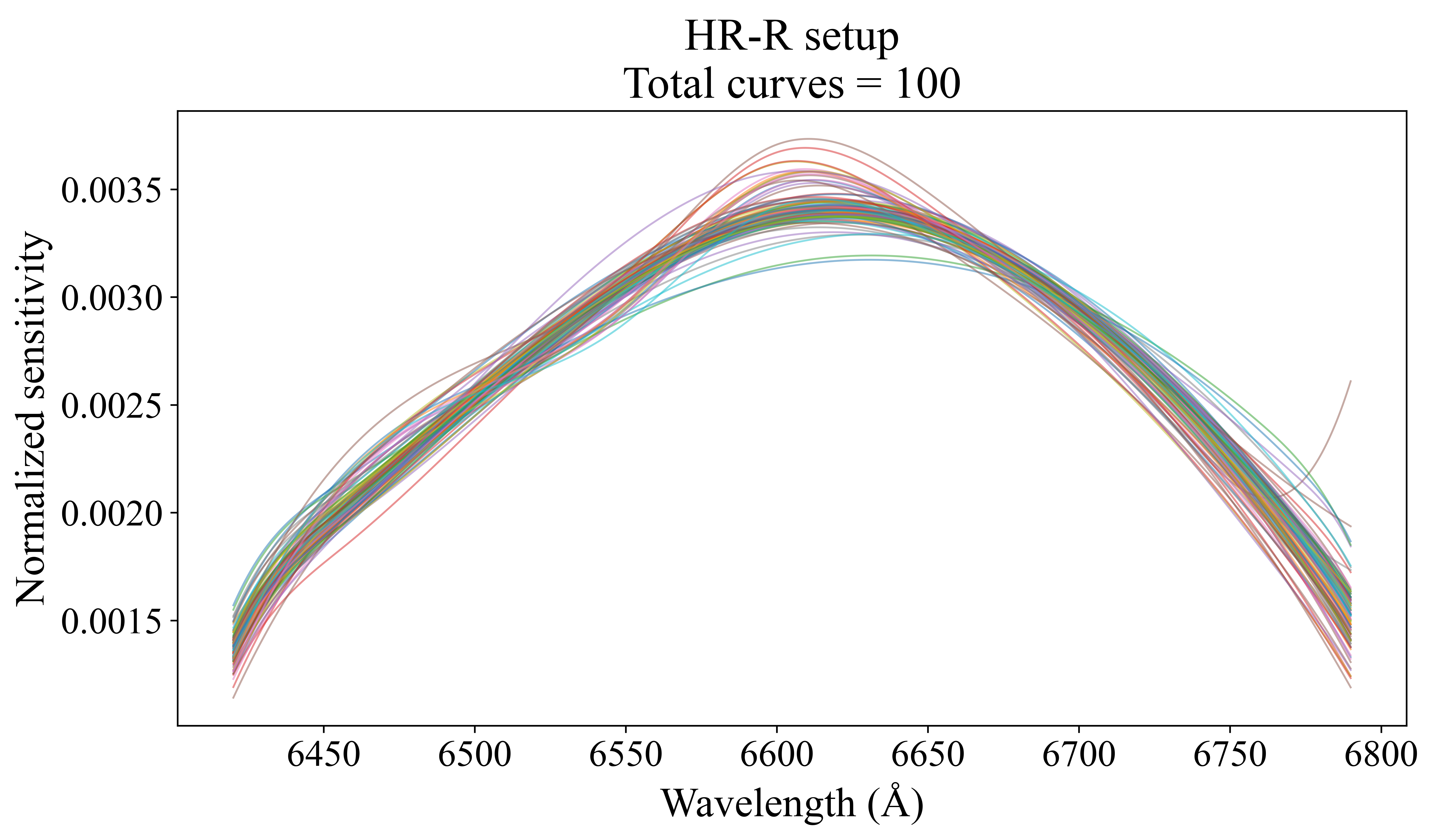}
\includegraphics[width=0.9\columnwidth,angle=0]    {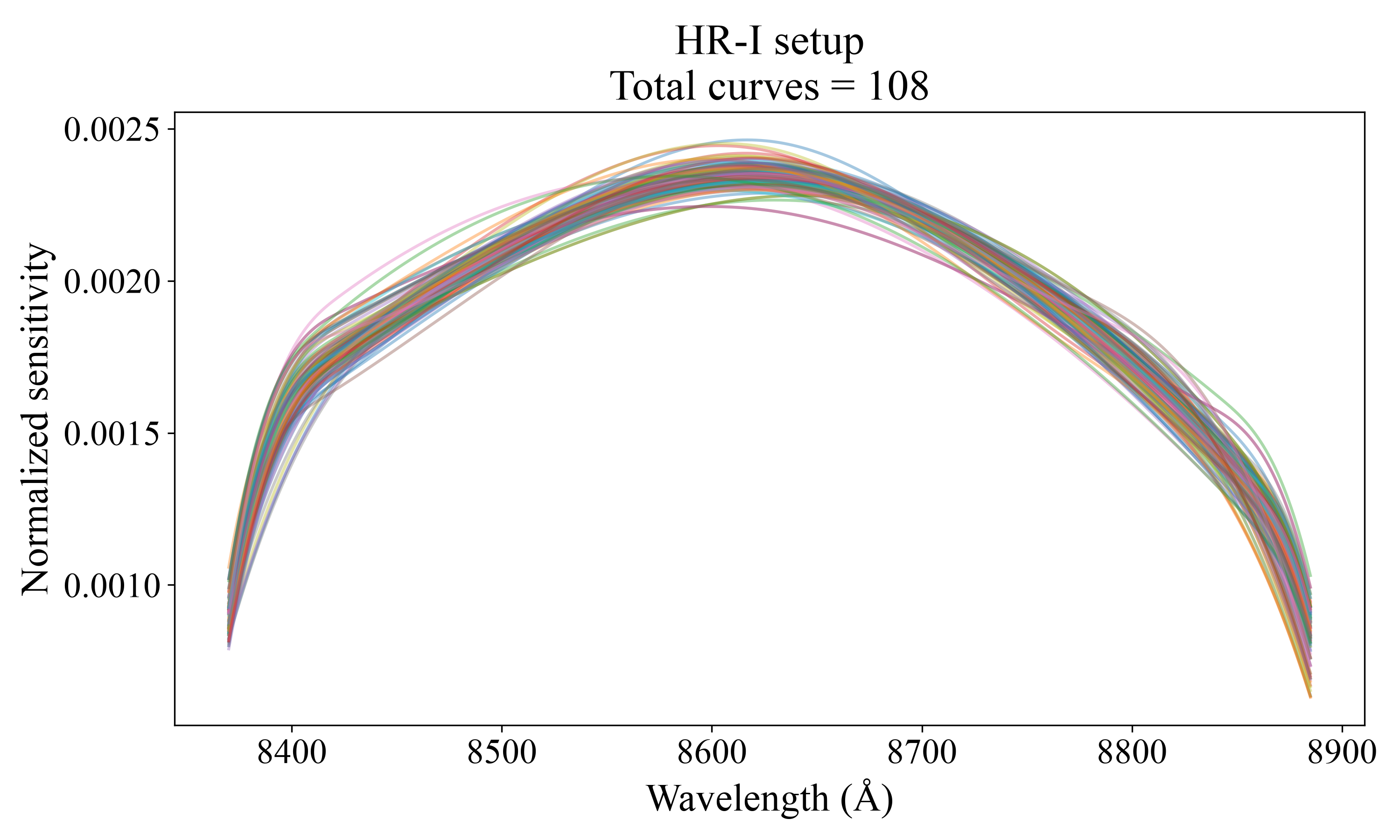}
\caption{Sensitivity functions obtained as  products  of the data reduction pipeline using the MegaraLcbStdStar recipe for the HR-R (top) and HR-I (bottom) spectral configurations. }
\label{sensitivity}
\end{center}
\end{figure}

\begin{figure*}[!ht]
\begin{center}
\includegraphics[width=0.8\columnwidth,angle=0]{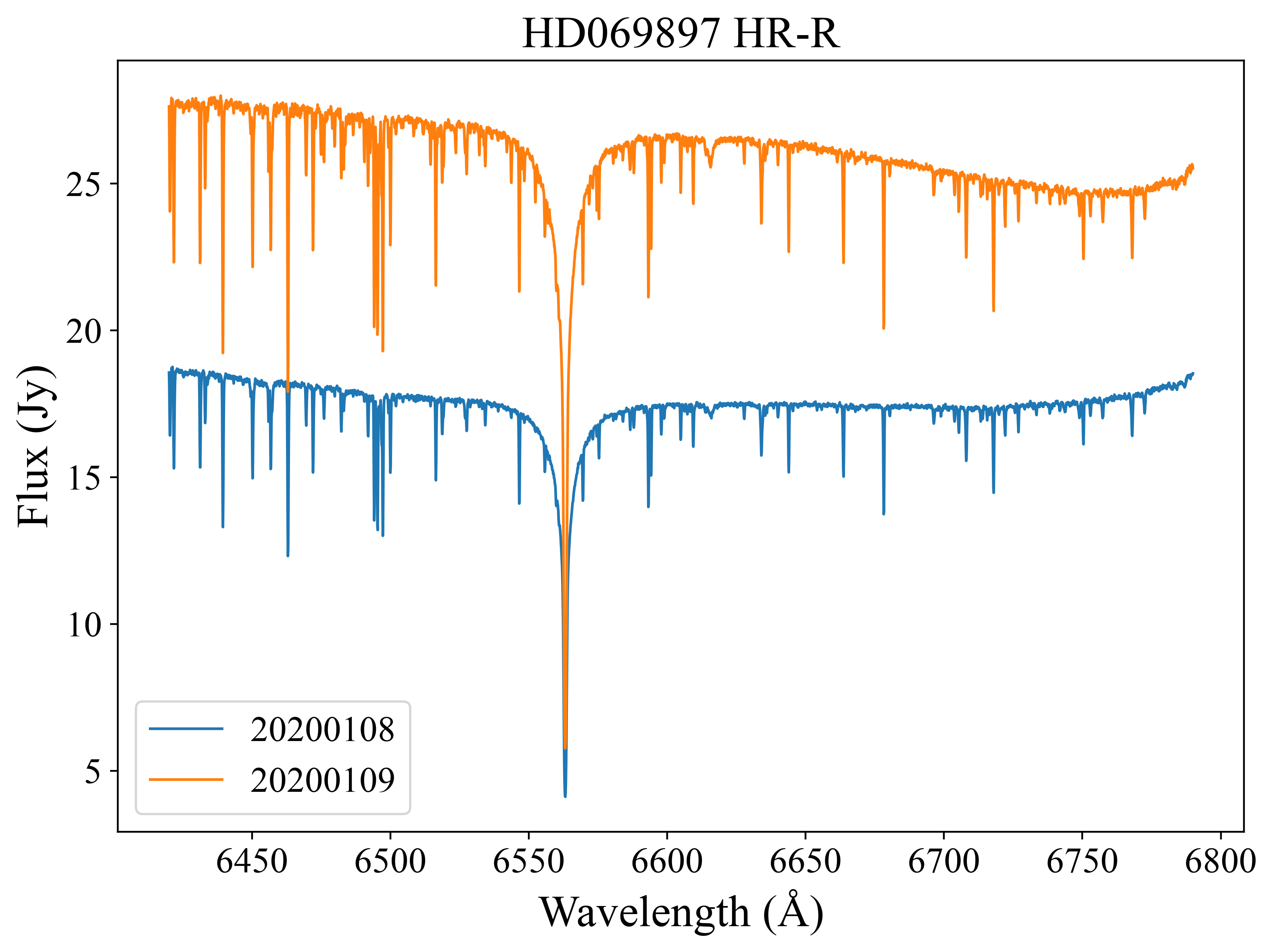}
\includegraphics[width=0.8\columnwidth,angle=0]{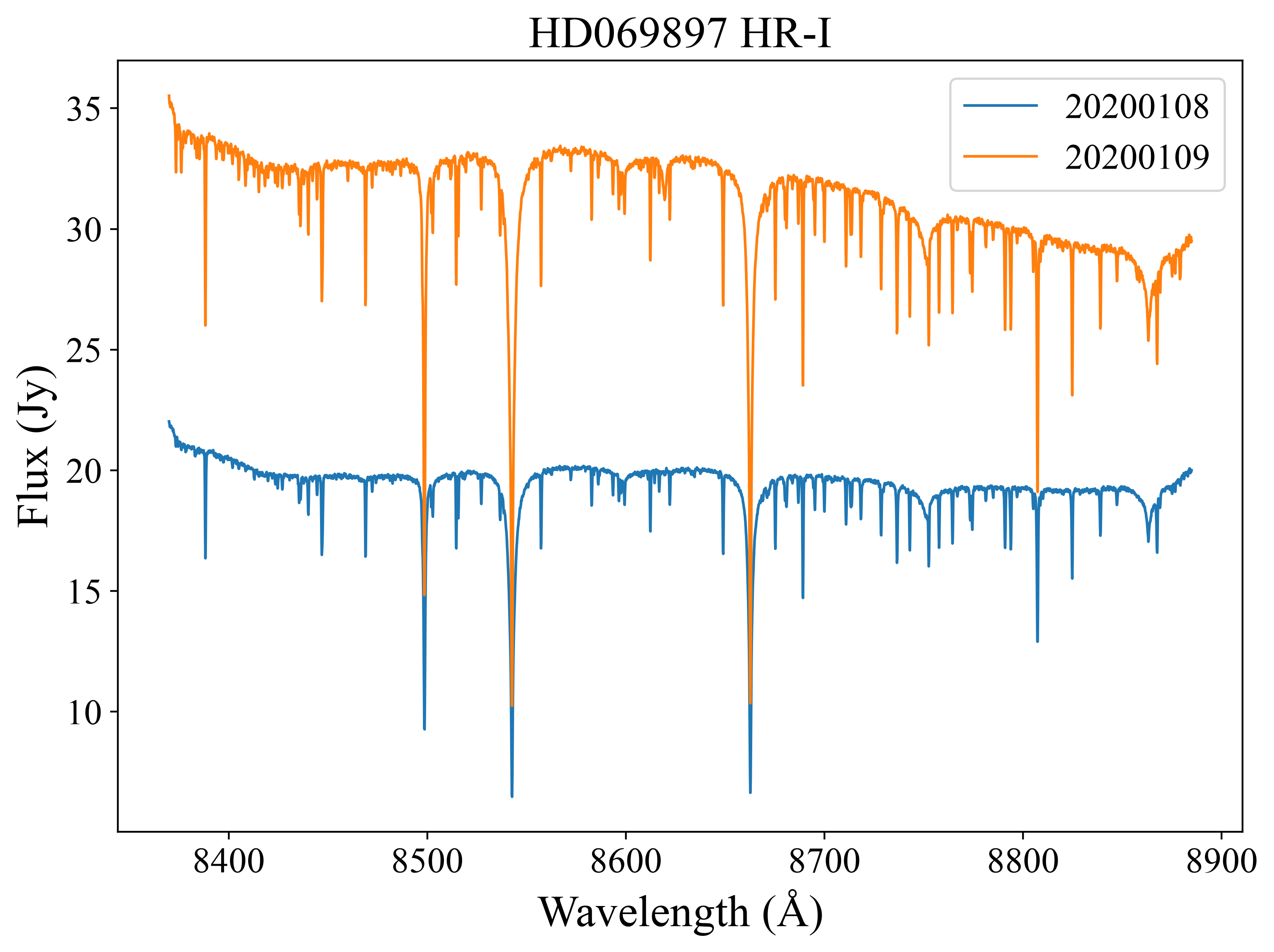}
\includegraphics[width=0.8\columnwidth,angle=0]{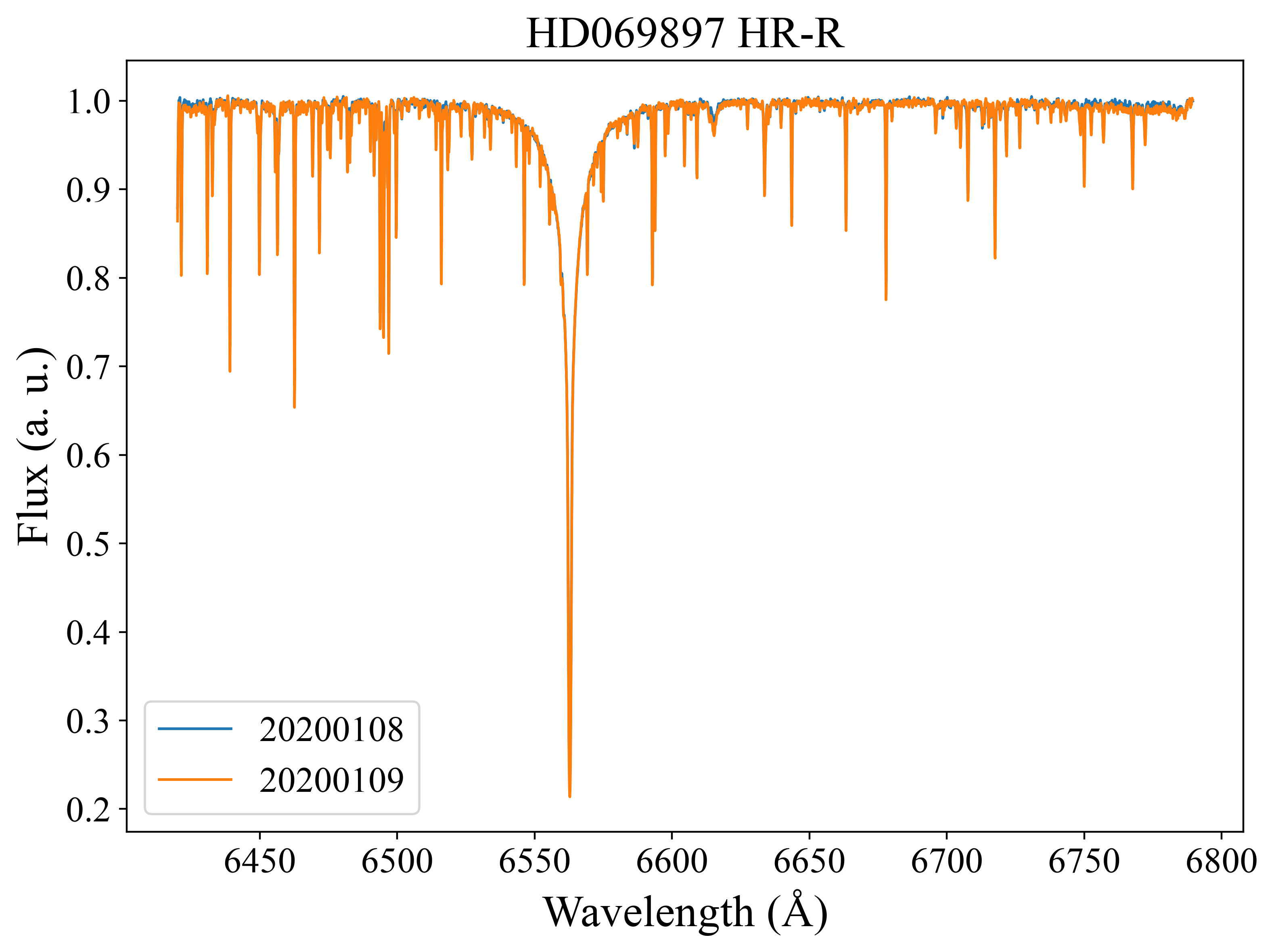}
\includegraphics[width=0.8\columnwidth,angle=0]{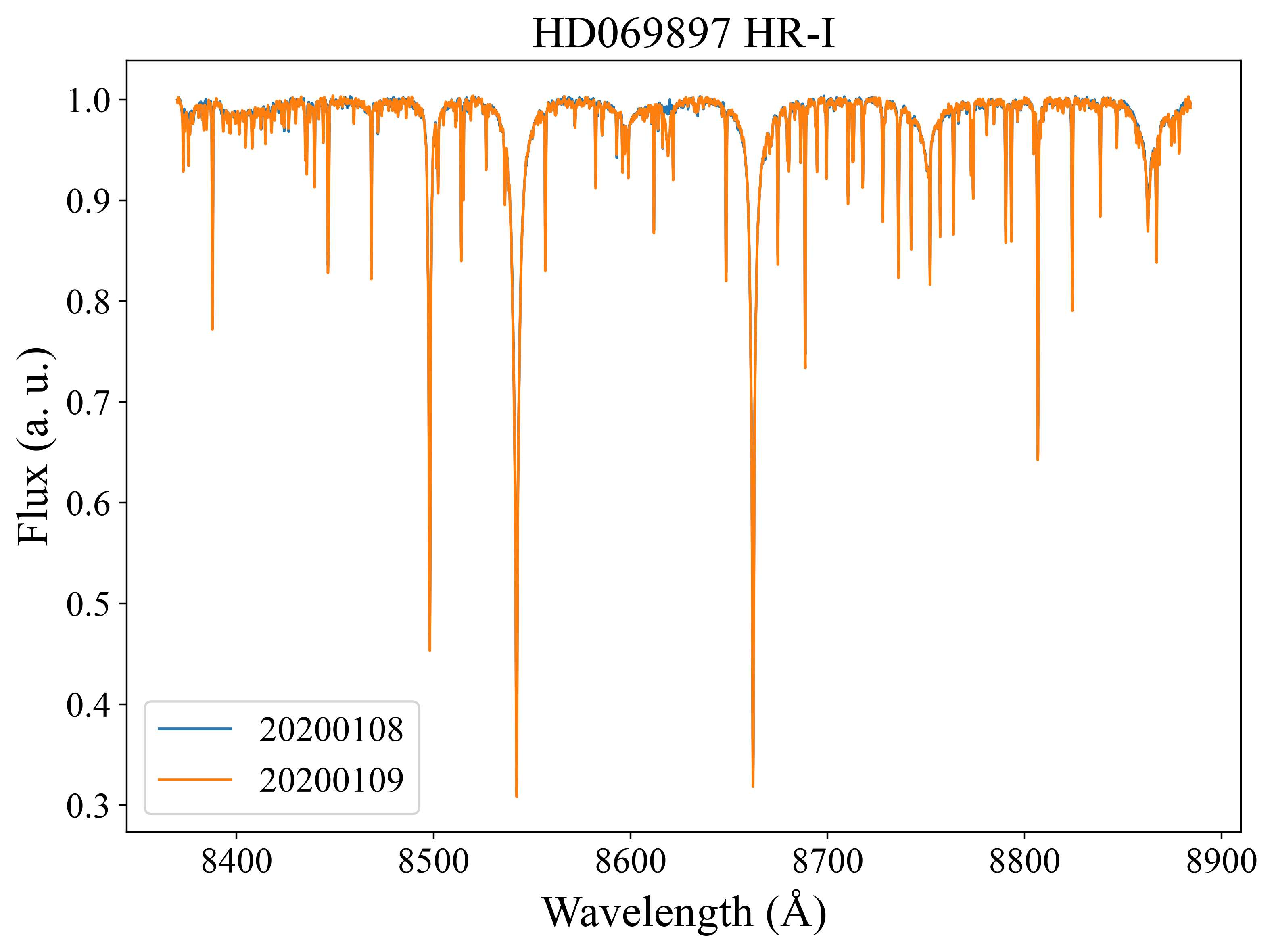}
\caption{Spectra in both setups of the star HD069897,  as indicated. The blue and yellow curves correspond to observations of the same star acquired on 8 January 2020 and 9 January 2020, respectively. The top panels show the flux-calibrated spectra provided in this data release, while the bottom panels present the  spectra rectified to the continuum and corrected for radial velocity, from \citet {molla23}.}
\label{stars_twice}
\end{center}
\end{figure*}

In both configurations, the largest calibration uncertainties occur near the spectral boundaries. To quantify this effect, we recomputed the RMS values after progressively trimming the wavelength coverage at both ends of the spectrum.
For HR-R, the RMS decreases from 3.46\% to 3.22\%, 3.10\%, 2.96\%, 2.83\%, 2.72\% and 2.62\% when excluding 5, 10, 15, 20, 25, and 30\,\AA\ from each edge of the spectrum, respectively. Similarly, for HR-I, the RMS decreases from 3.27\% to 3.08\%, 2.90\%, 2.73\%, 2.59\%, 2.47\% and 2.38\% for the same trimming intervals. The relatively modest reduction in RMS beyond 15\,\AA\ indicates that most of the degradation in the flux calibration is confined to a narrow wavelength region close to the spectral limits.
Adopting a wavelength range trimmed by 15\,\AA\ at both ends yields an estimate of the reproducibility of the relative flux calibration of approximately 3.0\% for HR-R and 2.7\% for HR-I. The internal consistency of 100 HR-R and 108  HR-I independently derived sensitivity functions indicates a relative flux-calibration variation of approximately 3\%.

 Fig. \ref{stars_twice} shows an example of two spectra of  HD069897  obtained for  both spectral configurations. Although differences in the relative flux are present between them, once the spectra are rectified to the continuum and corrected for radial velocity,  these differences do not affect  their use in determining the stellar parameters, see \citet{molla23}.

\subsection{Spectral Resolution}
An estimation of the variation in spectral resolution across the wavelength range, along the pseudo-slit, and among individual fibres was performed. The distribution of fibres in the IFU and along the pseudo-slit is such that fibres located near its centre are mapped to the central region of the pseudo-slit. For MEGASTAR observations, the GTC pointing accuracy is typically better than 0.7\,arcsec, and the support astronomers ensure that the target star is centred within the IFU. Consequently, the stellar signal is dominated by the central fibres, where the spectral resolution is both higher and more uniform. To quantify the variation in spectral resolution along the pseudo-slit, we considered three fibre selections: (i) all available fibres, (ii) the central region comprising fibres 200\,–\,400 and (iii) the innermost subset comprising fibres 250\,–\,350.
\begin{figure}[!ht]
\begin{center}
\includegraphics[width=0.9\columnwidth,angle=0]{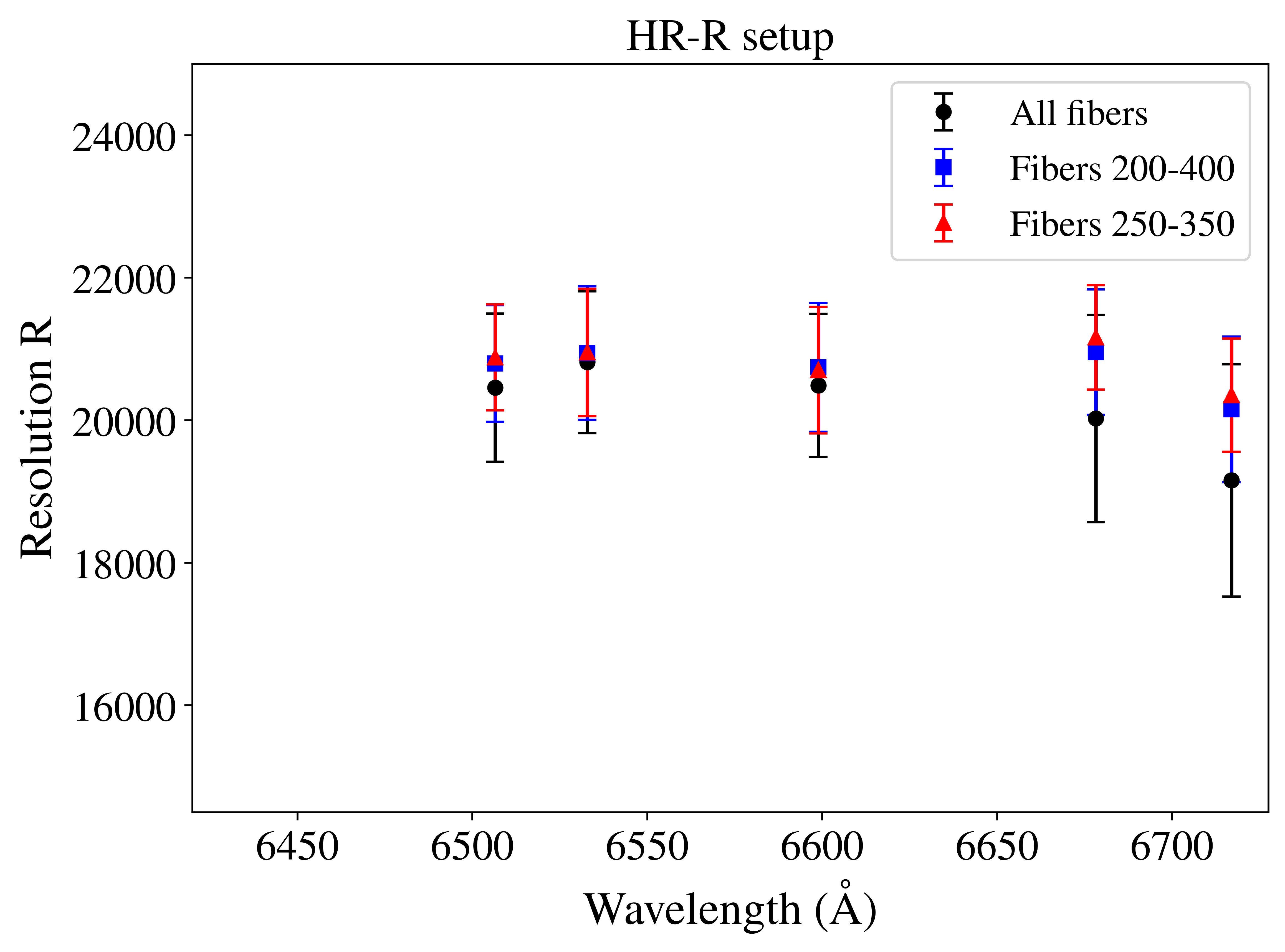}
\includegraphics[width=0.9\columnwidth,angle=0]{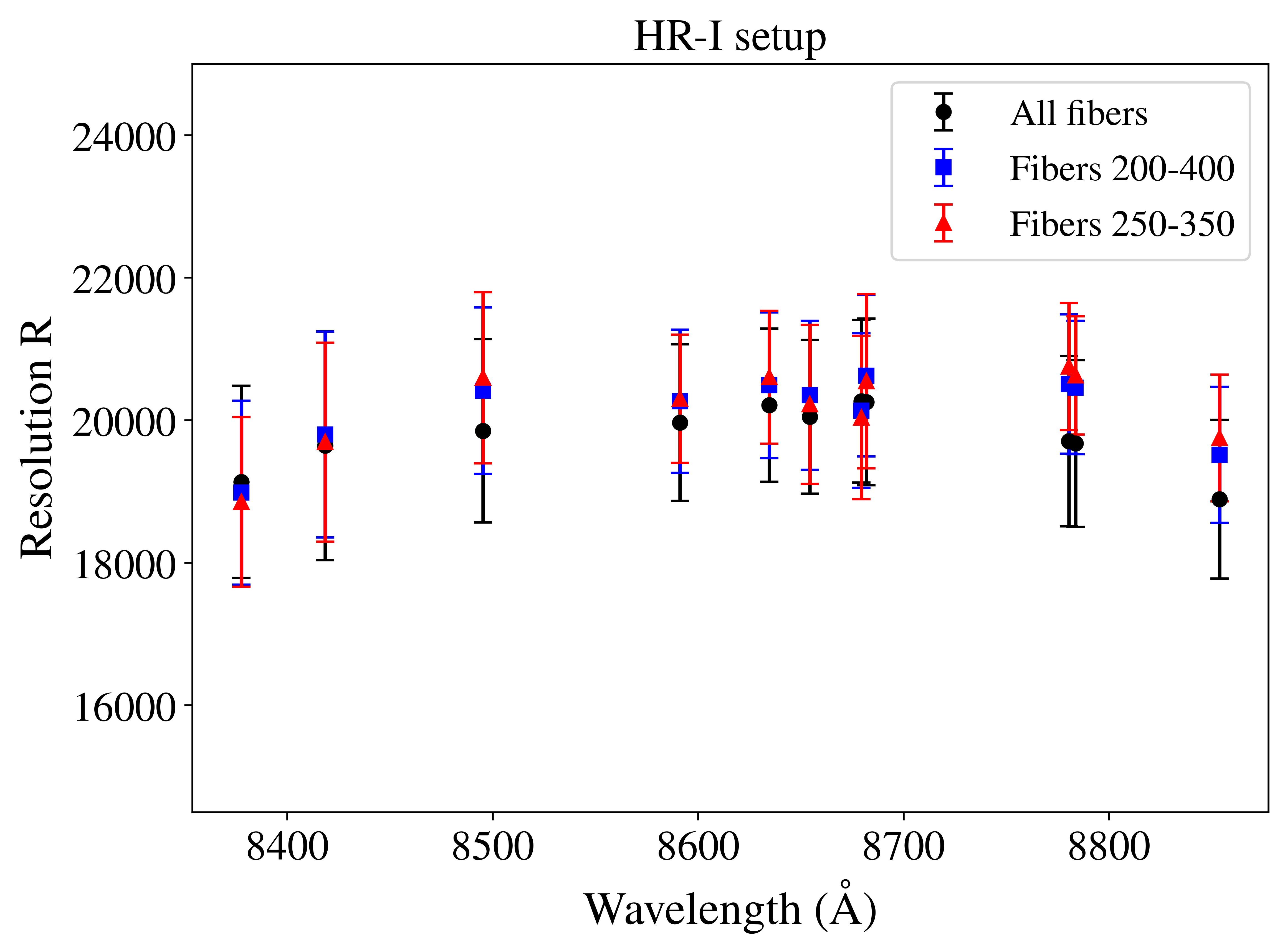}
\caption{Resolving power R = $\lambda/\Delta \lambda$ as a function of wavelength for HR-R and HR-I setups, as indicated. Error bars represent  $\pm$ 1$\sigma$. Black circles correspond to the mean resolving power obtained from all calibrated fibres for a given arc-lamp line. Blue squares and red triangles  correspond to the mean resolving power derived from fibres 200\,–\,400 and 250\,–\,350 respectively, the latter representing the central region of the pseudo-slit. 
}
\label{resolution}
\end{center}
\end{figure}
To evaluate the spectral resolution of the HR-R and HR-I configurations we used  the ThNe arc-lamp exposures obtained for wavelength calibration. For each identified arc line, the resolving power was computed as \mbox{R = $\lambda/\Delta \lambda$}, where $\Delta \lambda$ was obtained from the measured FWHM of the arc line and the local wavelength dispersion derived from the polynomial wavelength solution. For the \mbox{HR-R} setup five arc lines were identified across the spectral range 6420\,–\,6790\,\AA\,: 6506.5, 6532.9, 6598.9, 6678.3 and 6717.0 \AA\,. For the HR-I setup eleven arc lines, located at 8377.6, 8418.4, 8495.4, 8591.3, 8634.6, 8654.4, 8679.5, 8681.9, 8780.6, 8783.8, 8853.9\,\AA\, distributed across the spectral range 8370\,–\,8885\,\AA\, were used for the analysis. The resolving power was calculated independently for every valid calibration line in each fibre and subsequently analysed as a function of both wavelength and fibre position along the pseudo slit. The resulting measurements were used to characterise separately the wavelength dependence of the resolving power and its variation across the slit. 

The results for the HR-R and HR-I spectral configurations are shown in Fig. \ref{resolution}. For HR-R, the wavelength calibration provides a solution for 606 fibres. Considering the complete wavelength interval and all the pseudo-slit fibres, the mean resolving power increases from R\,=\,20456 at 6506.5\,\AA\ to a maximum value of R\,=\,20812 at 6532.9\,\AA\,, which represents a 1.7\% variation, and decreases to 19153 at 6717\,\AA\,, equivalent to about 8\%.  Both fibre subsets  (200\,–\,400 and 250\,–\,350) show very similar wavelength dependence, the maximum mean resolving power being R\,=\,20940 and R\,=\,20949, respectively. The central fibres provide a mean resolving power about 3\% higher than the average obtained from the full 606 set of fibres. The relative difference between the two central fibre selections is only 0.44\%.

\begin{figure}[!ht]
\begin{center}
\includegraphics[width=1\columnwidth,angle=0]{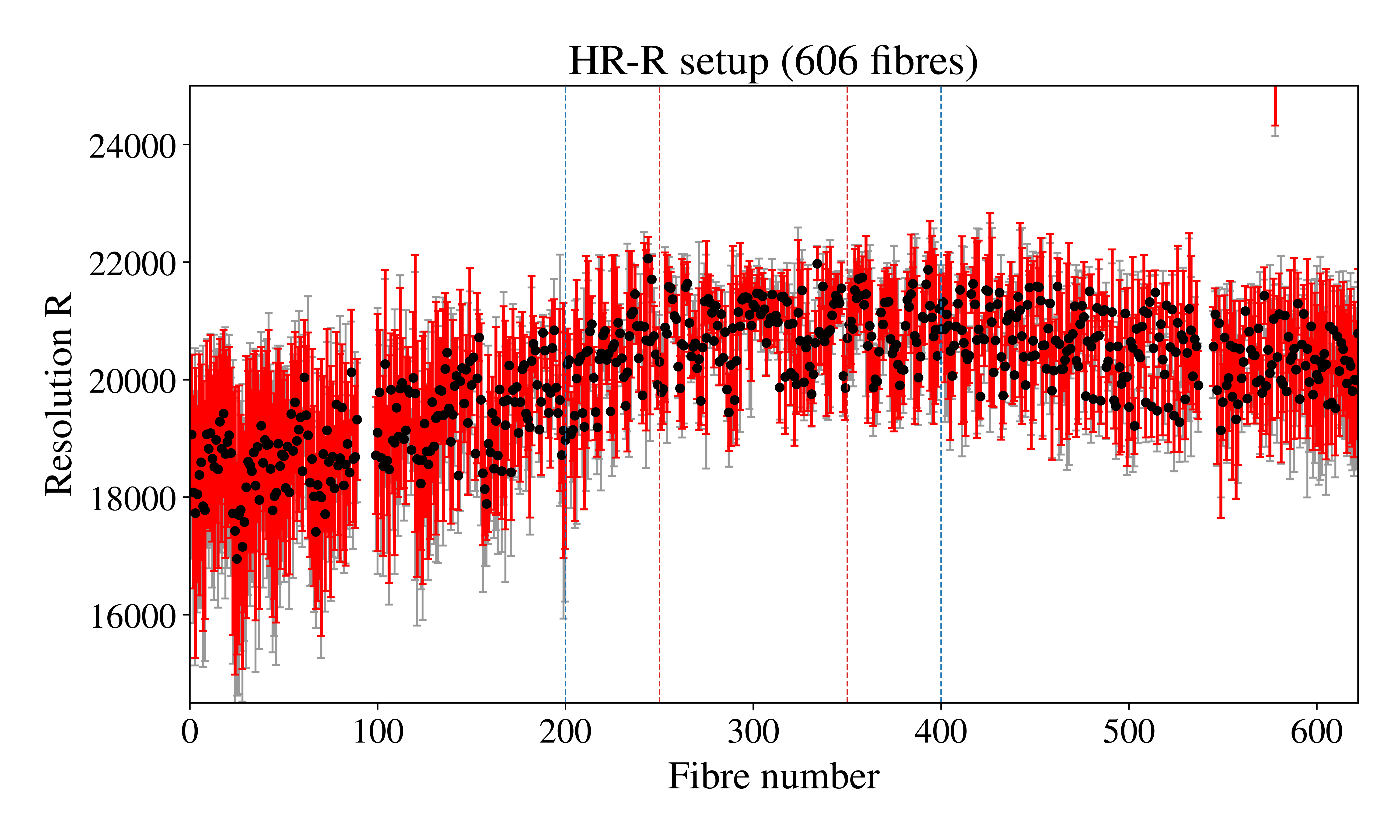}
\includegraphics[width=0.95\columnwidth,angle=0]{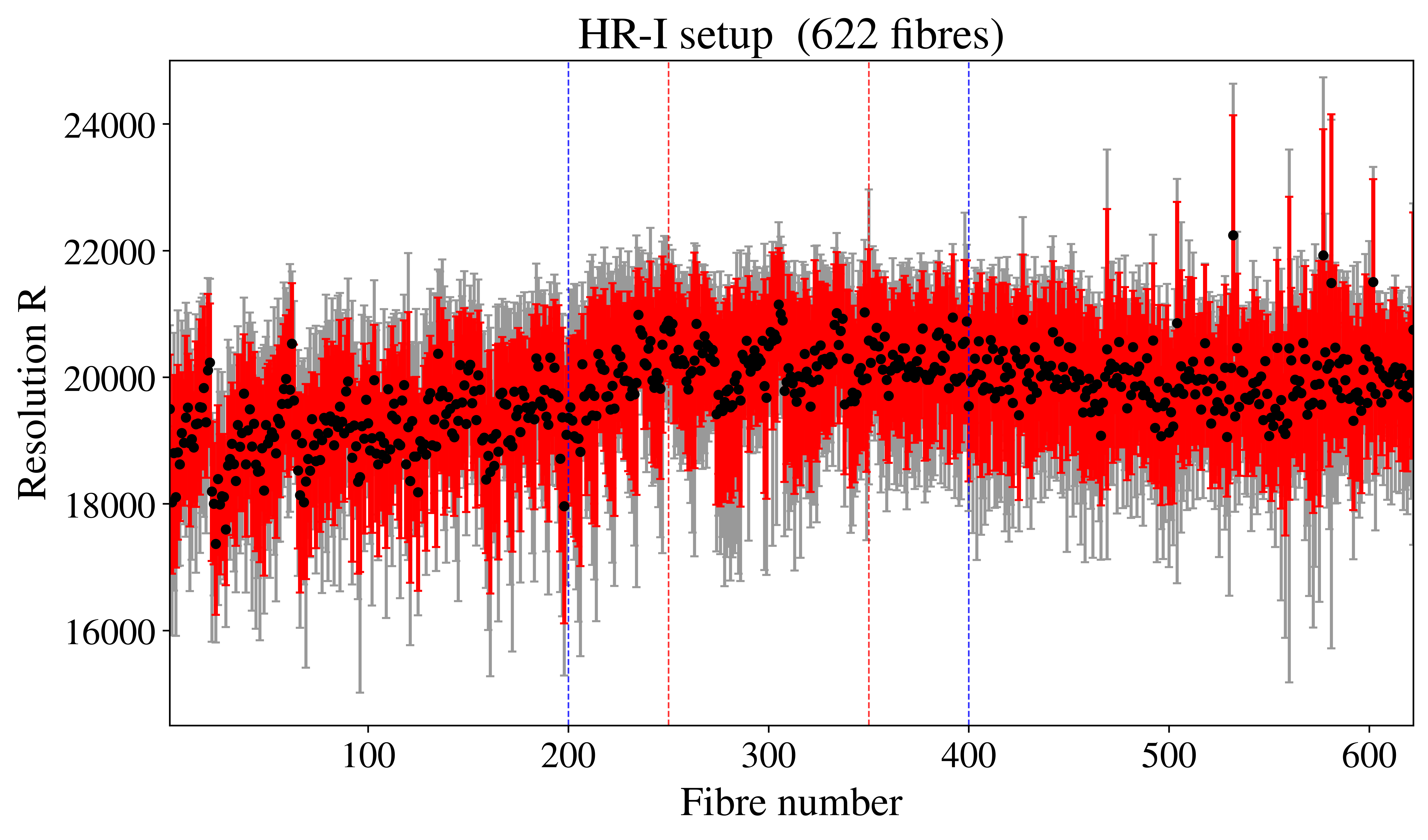}
\caption{Resolving power R\,=\,$\lambda/\Delta \lambda$ as a function of fibre number for the HR-R and HR-I setups, as indicated. For each fibre, the resolving power was computed independently for all available calibration lines and subsequently averaged. Black circles indicate the mean resolving power for each fibre, red error bars correspond to $\pm$\,1$\sigma$ and the grey bars span the minimum and maximum values measured. The vertical blue dashed lines mark the boundaries of the fibre subset 200\,–\,400 and the vertical dashed red lines indicate the subset 250\,–\,350.}
\label{resolution_fibres}
\end{center}
\end{figure}
For HR-I the mean resolving power considering the complete wavelength interval and the 622 pseudo-slit fibres is R\,=\,19785. The mean resolving power increases from R\,=\,19132 at  8377.6\,\AA\ to R\,=\,20266 at 8679.5\,\AA\, a variation of 5.6\%; the mean resolving power falls again to R\,=\,18890 at 8853.9\,\AA\,,  i.e. about 7\%.  For the fibre subsets (200\,–\,400 and 250\,–\,350) the mean resolving power obtained was R\,=\,20139 and R\,=\,20180, respectively. The fibre selections show very similar wavelength dependences differing primarily by an approximately constant offset in resolving power. The central fibres provide a resolving power about 2\% higher than the average obtained from  the full pseudo-slit. Furthermore, the relative difference between the two central fibre selections is only 0.20\%, indicating that the central region of the pseudo-slit is highly uniform in terms of resolving power.

The fibre-to-fibre scatter was quantified as the standard deviation of the distribution of the fibre-mean resolving powers as shown in Fig. \ref{resolution_fibres}. For HR-R, the mean resolving power obtained from all calibrated fibres is R\,=\,20125, with a fibre-to-fibre standard deviation of 1016, corresponding to a relative scatter of 5.1\%. Restricting the analysis to fibres 200\,–\,400 and 250\,–\,350 yields mean resolving powers of R\,=\,20716 and R\,=\,20808, respectively, with corresponding fibre-to-fibre standard deviations of 606 (3.0\%) and 599 (2.8\%). The relative difference between the two central fibre selections is only 0.44\%. For HR-I, the mean resolving power obtained from all 622 fibres is R\,=\,19759, with a fibre-to-fibre standard deviation of 634 (3.2\%). Restricting the analysis to fibres 200\,–\,400 and 250\,–\,350 yields mean resolving powers of R\,=\,20143 and R\,=\,20196, respectively, with corresponding fibre-to-fibre standard deviations of 428 (2.1\%) and 407 (2.0\%). The relative difference between the two central fibre selections is only 0.20\%.
These results indicate that the apparent variation among fibres is dominated by a smooth center-to-edge trend across the pseudo-slit rather than by random fibre-to-fibre fluctuations. The central region of the pseudo-slit provides both the highest resolving power and the greatest uniformity. Furthermore, the residual fibre-to-fibre scatter in the central region is only about 2\,–\,3\% for both setups, substantially smaller than the wavelength-dependent variation of the resolving power discussed above.

\end{appendix}
\end{document}